\newif\ifprinter
\printertrue
\documentstyle [preprint,revtex,epsf,eqsecnum]{aps}
\ifpreprintsty
    \advance\textwidth by 0.395in
    \def\prdbrk#1{}
    
    \makeatletter		% override revtex's mangled \footnote
    \def\footnote{\@ifnextchar[{\@xfootnote}{\stepcounter
       {\@mpfn}\xdef\@thefnmark{\thempfn}\@footnotemark\@footnotetext}}
    \makeatother
\else
    \def\prdbrk#1{\nonumber\\&&#1{}}
    
\fi
\makeatletter		% fix revtex's mangled footnotes & figures.
\def\footnotesize{\@setsize\footnotesize{9pt}\xpt\@xpt
\abovedisplayskip 8pt plus2pt minus4pt
\belowdisplayskip \abovedisplayskip
\abovedisplayshortskip \z@ plus1pt
\belowdisplayshortskip 4pt plus2pt minus2pt
\def\@listi{\topsep 4pt plus 2pt minus 2pt
\parsep 2pt plus 1pt minus 1pt \itemsep \parsep}}
\def\footnote{\@ifnextchar[{\@xfootnote}{\stepcounter{\@mpfn}%
    \xdef\@thefnmark{\thempfn}\@footnotemark\columnwidth=\hsize\@footnotetext}}
\skip\footins 16pt plus 4pt minus 4pt
\def\fnum@figure{Figure \thefigure.}
\def\figure{\@float{figure}}
\let\endfigure\end@float
\@namedef{figure*}{\@dblfloat{figure}}
\@namedef{endfigure*}{\end@dblfloat}
\makeatother
\begin {document}
\draft
\preprint {UW/PT-92-04}
%\preprint {DOE/ER/40614-19}
\begin {title}
    {%
    Quantum Field Theory in Spaces with Closed Time-Like Curves
    }%
\end {title}

\author {David G. Boulware}

\begin {instit}
    {%
    Department of Physics FM-15, \\
    University of Washington, \\
    Seattle, Washington 98195
    }%
\end {instit}
\receipt{}
\begin {abstract}
Gott spacetime has closed timelike curves, but no locally anomalous
stress-energy.
A complete orthonormal set of eigenfunctions of the wave
operator is found
in the special case of a spacetime in which the
total deficit angle is $2\pi$.
A scalar quantum field theory is constructed using these eigenfunctions.
The resultant interacting quantum field theory is not unitary because
the field operators can create real, on-shell, particles in the acausal
region.
These particles propagate for finite proper time accumulating an arbitrary
phase before being annihilated at the same spacetime point as that at which
they were created.
As a result, the effective potential within the acausal region is complex,
and probability is not conserved.
The stress tensor of the scalar field is evaluated in the neighborhood of the
Cauchy horizon; in the case of a sufficiently small Compton
wavelength of the field,
the stress tensor is regular and cannot prevent the formation of the Cauchy
horizon.

\end {abstract}
\vskip -5ex plus 5ex
\ifpreprintsty\else
\pacs{3.70.+k,4.20.Cv}
\fi

\ifpreprintsty
{
\centerline{\small PREPARED FOR THE U.S. DEPARTMENT OF ENERGY}

%\medskip

\baselineskip 8truept plus0.2pt minus0.2pt

\noindent
{\tiny  This report was prepared as an account of work
sponsored by the United States Government.  Neither the United States nor
the United States Department of Energy, nor any of their employees, nor
any of their contractors, subcontractors, or their employees, makes any
warranty, express or implied, or assumes any legal liability or responsibility
for the product or process disclosed, or represents that its use would not
infringe privately-owned rights.
By acceptance of this article, the publisher and/or recipient acknowledges
the U.S. Government's right to retain a nonexclusive, royalty-free license
in and to any copyright
covering this paper.}

}

\fi
\ifpreprintsty
\pagebreak
\fi
\narrowtext
\setcounter{page}{1}
\section{Introduction}
\label{Introduction}

Spacetimes with closed timelike lines have generally been considered
unphysical \cite{Hawking}
because of logical paradoxes, the lack of a well posed
Cauchy problem, or the sense that they are obviously wrong.
Prompted by the work of Morris
{\it et al.\/}\ \cite{Thorne:wormhole:lett,Thorne:wormhole:ajp},
there has recently been an
extensive reexamination of the question.  The conclusion of this reexamination
is that it is not trivial to decide whether closed timelike curves are
physically allowed.
Indeed, spacetimes with closed timelike curves exist, and solutions to field
equations on these spaces exist.
These solutions are complete on some spacelike surfaces in at least some
acausal spacetimes \cite{Friedman:cauchy:lett,Friedman:cauchy}.
Although the causality properties of these spacetimes are unfamiliar, they do
not appear to be self-contradictory, and, if one is prepared to consider them
at all, one must address the question of their acceptability in other terms.

The original wormhole spacetimes of Morris
{\it et al.\/}\ \cite{Thorne:wormhole:lett}
require
that the stress-energy which supports the wormhole fail to satisfy the
positive energy condition.
Although the matrix elements of the
stress-energy of a quantum field do not in general satisfy the positive energy
condition, their volume integrals over distances large compared with the wave
lengths associated with the field excitations are in general positive, and
it is still somewhat problematic to
reconcile the existence of the wormholes with the stability of matter.
On the other hand, Gott\footnote{See also Deser {\it et al.\/}\ \cite{DJt:3d}.}
\cite{Gott} has pointed out that spacetimes with
two relatively
moving infinitely long straight strings can possess closed timelike curves.
These spacetimes are vacuum spacetimes except for conical singularities
at the strings, and each string alone is a physically acceptable solution to
Einstein's equations.
Although
Carroll {\it et al.\/}\ \cite{Guth:momentum}
and
Deser {\it et al.\/}\ \cite{DJt:momentum}
have pointed out
that such spacetimes cannot arise from a spacetime which initially contains
no closed timelike curves and has a positive definite total energy, the
Gott spacetime itself does not have any local properties which are
physically unacceptable.
The existence of the acausal region in which future directed non-spacelike
curves can intersect each other is the only peculiarity of the spacetime.

These considerations suggest that it is worth studying the properties of
Gott space in more detail.
A point mass in $2 + 1$ dimensions produces a spacetime which is everywhere
flat.
Coordinates may be chosen in which the metric is the flat
Minkowski metric,
but with a wedge removed from the space, and the points along the edges of the
wedge identified.
The resultant cone has a singularity at its tip where the
point mass which produces the space is located.
The circumference of a circle of radius $r$ is $(2\pi - \theta)r$, where
$\theta$ is the deficit angle of the cone and is a measure of the
mass \cite{DJt:3d}.
The Gott spacetime is generated by two such point masses moving relative to
one another.
The scalar wave equation is particularly easy to analyze in the special
limiting case in which the deficit angles of the two points are both $\pi$.
(That space is open, whereas if the sum of the deficit angles were greater
than $2\pi$ the space would be closed \cite{Deser:2dim}.)
In particular, a scalar quantum field may be constructed on the space using
a path integral to calculate the propagators.
The resultant free field theory appears to be fully acceptable; however, an
interacting field theory is not unitary.

The procedure for constructing the field theory is fairly straightforward.
Coordinates may be chosen in which the metric is the Minkowski metric,
but with the edges of the removed wedges identified with additional boosts.
Because of the boosts, the point (in $2+1$ dimensions) or string
(in 3+1 dimensions) singularities are moving.
Despite the unusual boundary conditions, it is possible to solve the scalar
wave equation on the spacetime.
A complete orthonormal set of functions which are
eigenfunctions of the wave operator is exhibited in Section \ref{eigen}.
Using these eigenfunctions, the functional integral which defines,
in causal spacetimes, the time-ordered matrix elements of the field
operators in the vacuum will be evaluated in Section \ref{Green}.
The resulting matrix elements appear as an infinite sum of terms
corresponding
to the various possible winding numbers of paths around the singularities.
Alternatively, each term may be viewed as
corresponding to a given image under (boosted)
reflection in the boundaries.
In this acausal spacetime the individual terms are
precisely the functions which one would
naively write down with, however, each term in the sum being separately time
ordered according to whether the field point is in the past or future of the
source point image.
As a result, when both the points of
the matrix element of a pair of field operators (or Green's function)
are in the acausal
region, the points may be connected by a future directed timelike curve for
some winding numbers and by a past directed timelike curve for other winding
numbers.
Because of this, the functional integral result for
$\langle (\phi(x) \phi(x'))_+\rangle$ cannot be regarded as a matrix element
in which the field associated with the earlier time lies to the right of the
field associated with the later time.
When both points are in the acausal region, each point is both in the future
and in the past of the other point.
The functional integral defines the matrix element by having either field
create
positive energy excitations which travel along future directed timelike curves
to be annihilated by the other field.

To put this another way, time-ordered products cannot be constructed because
there is not a well defined time-ordering for pairs of points in the acausal
region.
However, the propagation of particles (or waves) is well ordered in that they
propagate forward in time.
(These spacetimes do have a well defined direction of time.)

The complete set of eigenfunctions of the wave operator then
provide a complete set of solutions of the wave equation.
These are complete on a given spacelike surface in the causal region,
and that completeness constitutes a special case of the theorem proven by
Morris and Friedman \cite{Friedman:cauchy}:
It is possible to arbitrarily specify the positive frequency field on an
initial spacelike surface which lies entirely outside the acausal region;
this uniquely determines the positive frequency field throughout the
spacetime.

If a Hermitian product of field operators acts
at the same spacetime point
in a causal spacetime, its vacuum matrix element must be real
because it can only create virtual particles which are reabsorbed at
the same point.
If the point is in the causal region of an acausal spacetime, the same result
holds, but in the acausal region one field can create an excitation which
propagates forward in time, accumulating an arbitrary positive phase, to be
annihilated by another field acting at the same spacetime point.
Since no negative phases can be produced (particles always propagate forward
in time), the matrix element of the product of the field operators in the
vacuum will not in general be real.
The fact that
matrix elements of `physical' operators such as current densities and
stress-energy tensors no longer possess the expected reality properties in the
acausal region can be ignored in a non-interacting theory because
these operators are uncoupled in that theory.
The lack of Hermiticity appears only in multiplicative factors of the
matrix elements which are canceled in the renormalization process.

The situation is quite different for an interacting field.
The equation for the field itself contains products of more than one
field.
In a causal spacetime and in the causal region of an acausal spacetime, the
vacuum expectation value of the interaction term becomes a real effective
potential in which the excitations propagate.
In the acausal region of an acausal spacetime, the vacuum expectation value
gives an effective potential which is complex due to the phases of the
particles which are created and reabsorbed.
This potential does not yield unitary propagation, and the lack of unitarity is
not associated with the creation of particles which appear at future infinity
$I^+$.
It is
associated with the real creation of particles which are reabsorbed by the
same interaction.
That is, it is associated with what happens within the acausal region.
In the case of an interacting field,
complete information about what happens there is not
available on spacelike surfaces restricted to the causal region, and
the data which establishes the state of the system in the acausal region
cannot be given on such surfaces.
The first order corrections due to the complex potential produce probabilities
which may be greater than one, thereby violating unitarity in any sense.

There has been some discussion of the response of
the metric to the stress tensor induced by the existence of an acausal
region \cite{Thorne:vacuum,Hawking:chronology-protection}.
In the model discussed here, since the propagator is calculated exactly,
it is straightforward to calculate the matrix element of the stress-energy
tensor.
In the causal region it has the correct Hermiticity, and, after it has been
renormalized, it is regular everywhere in the causal region
but becomes singular as the Cauchy horizon is approached.
This singularity is of order $1/[x^{\pm}\ln^2(x^+x^-)]$,
where $x^{\pm} = 0$ defines the Cauchy horizon.
When this is inserted into Einstein's equations, the resultant
metric is of the form
\begin{equation}
ds^2 = e^{2\psi(-x^+x^-)}dx^+dx^- + e^{2\phi(-x^+x^-)}dy^2
\end{equation}
where $e^{2\psi} \rightarrow (-x^+x^-)^{2 C\ln\ln(Y^2_0/(-x^+x^-))}$ as
$ -x^+x^- \rightarrow 0^+$.
For zero mass fields,
this singularity is fairly weak, and it is not clear what a simultaneous
solution of Einstein's equations and the field equations would yield as a
self-consistent solution.
Nor is it clear how this relates to Hawking's chronology protection
conjecture \cite{Hawking:chronology-protection}.
However, for non-zero mass fields and for sufficiently small relative
rapidity of the point masses relative to their separation,
there is no singularity of the stress-energy and, therefore,
no mechanism for chronology protection.

Hartle \cite{Hartle} has studied the problem of unitarity in acausal spacetimes
using the decoherence approach of Gell-Mann and Hartle \cite{Gell-Mann}.
He too finds that there is a lack of unitarity, but for cases where there
is an acausal region to the future of the spacelike surfaces on which
measurements are made, in addition to the case
where the acausal region lies between the
surfaces.
The procedure followed in this work exhibits a lack of unitarity only when the
acausal region lies between the surfaces on which measurements are made.
This seems more reasonable in that our ability to make consistent
measurements now should not be compromised by the behavior of the system at
points which are in the future of the entire region in which the measurements
are made.

The order of presentation is as follows:
The following section consists of a brief outline of Gott space, establishing
the conventions used here.
The causal properties of the space are discussed in Section \ref{acausal},
followed by, in Section \ref{eigen}, a derivation of the complete
orthonormal set of eigenfunctions of the wave operator on the space.
The Green's function and operator products derived in Section \ref{Green} are
used in  Section \ref{stress} to derive the properties of
the matrix elements of the stress tensor and the operator $\phi^2$.
These results are used in Section \ref{interacting} to discuss the properties
of
the interacting field, and exhibit the lack of unitarity of the interacting
field.
The effect of the quantum field on the metric is discussed in the last
section.

\section{Gott Space}
\label{Gott}

The general matter-free solution to Einstein's equations in $2+1$ dimensions is
everywhere flat, and the solution with point masses is flat except for conical
singularities at the locations of the masses \cite{DJt:3d}.
The masses are proportional to the deficit angles at the singularities.
A $3+1$ dimensional spacetime with infinitely long straight strings
running parallel to the $z$ axis is also
flat with conical singularities along the strings.

Gott \cite{Gott} pointed out that two relatively
moving point masses in $2+1$ dimensions
(or two strings in $3+1$ dimensions) produce a spacetime which has closed
timelike lines.
He considered the physically realistic case in which the deficit angles of the
masses are small, and, as a consequence, the relative velocity of the masses
must be large in order to produce an acausal spacetime.
The space is open only if the sum of the deficit angles associated with the
masses is less than or equal to $2\pi$ \cite{DJt:3d}.
Only the limiting case of two equal masses with deficit angle $\pi$ will
be considered in this work; in that case it is easy to
obtain explicit solutions to the various equations.
In general, only the expressions in $2+1$ dimensions will be presented;
the straightforward
generalizations to $3+1$ dimensions will be given when required for
comparison purposes.

The spacetime produced by a single point mass with deficit angle $\pi$ may be
described by a half plane, as shown in Fig.\ \ref{fig:1mass},
with the edge identification $x \doteq - x $ at $y = 0\,.$
\ifprinter
\ifpreprintsty
\begin {figure}[htbp]%			Figure 1
    \vbox to 2.3in{
	\vss
	%\vskip -0.75in
	\leavevmode {\hfill \epsfbox {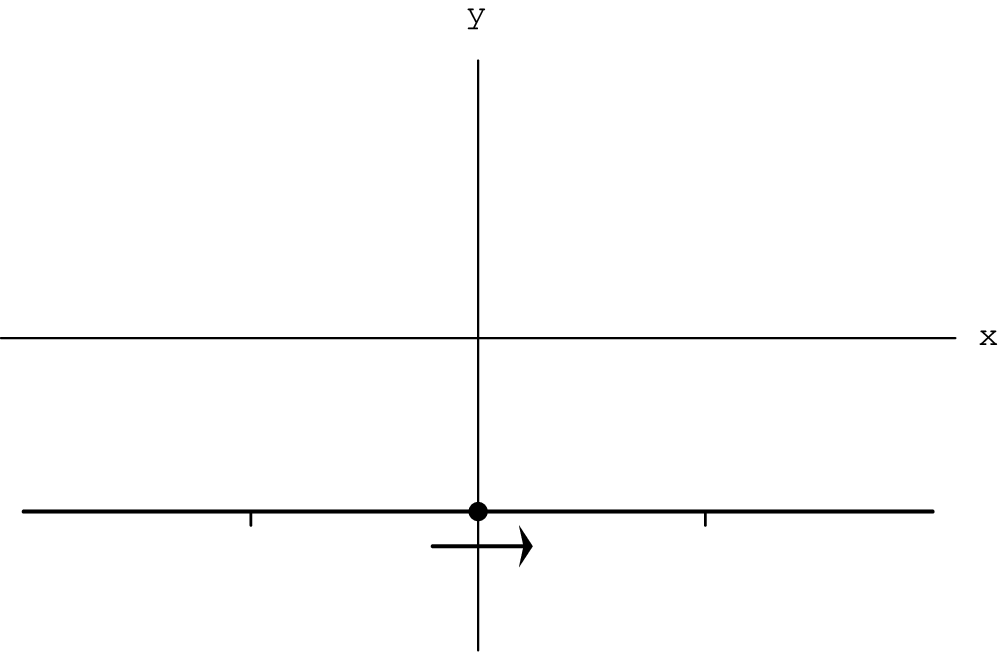} \hfill}
	%\vskip -0.75in
	\vss
	}
    \caption
	{%
	\advance\baselineskip by -8pt
	The space for a single point mass of deficit angle $\pi$. The space
	consists of the region above the line with the tick marks.
	The ticks denote identified points on the right and left sides of the
	line.
	}%
    \label {fig:1mass}
\end {figure}
\fi
\fi
\ifprinter
\ifpreprintsty
\begin {figure}[htbp]%			Figure 2
    \vbox to 2.3in{
	\vss
	%\vskip -0.75in
	\epsfxsize=2.5in
	\leavevmode {\hfill \epsfbox {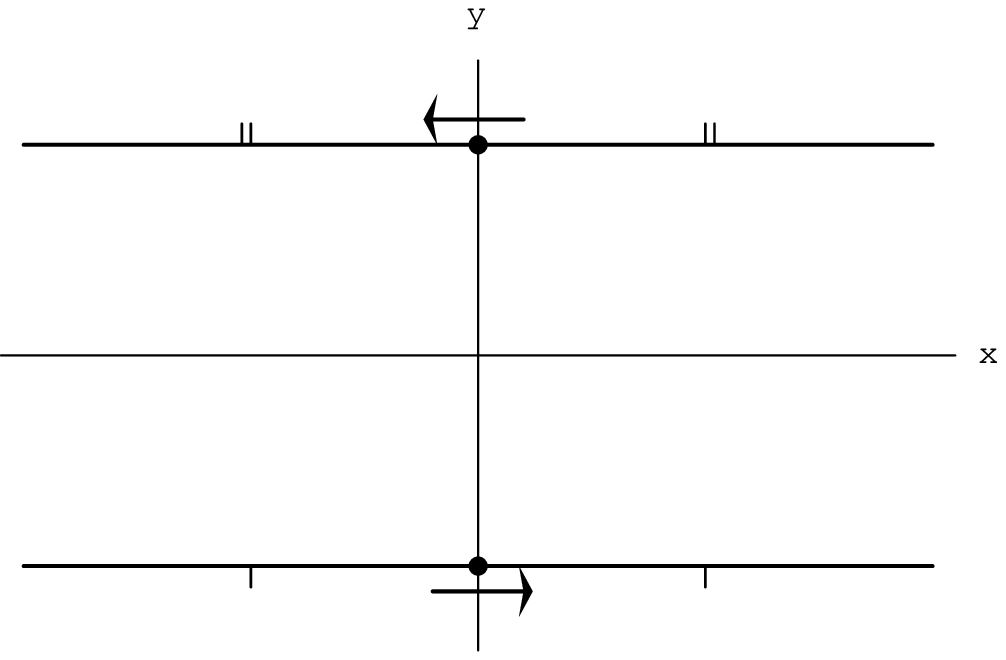} \hfill}
	%\vskip -0.75in
	\vss
	}
    \caption
	{%
	\advance\baselineskip by -8pt
	The space for two point masses each with deficit angle $\pi$. The space
	consists of the region between the lines with the single and double
	tick marks respectively denoting the identified lines just as in
	Fig.\ \ref{fig:1mass}.
	}%
    \label {fig:2mass}
\end {figure}
\fi
\fi
This physical space may be extended to a covering space
consisting of the entire $x$--$y$ plane, and an arbitrary function on the
physical space may be represented by a function on the covering space
which satisfies the condition
\begin{equation}
    f(x, y, t) = f( -x, -y, t)\,.
\end{equation}
Such a function is uniquely determined by the function on the physical
space.
An arbitrary  $C^{\infty}$ function $\phi(x, y, t )$
on the covering space yields a $C^{\infty}$
function on the physical space provided that $\phi(x,0,t) = \phi( -x, 0,t)$.
The eigenfunctions of the wave operator on the physical space are
eigenfunctions of the wave operator on the covering space
which satisfy this continuity condition.
Given $\phi(x, y, t)$, an arbitrary eigenfunction of the wave operator on the
covering space, an eigenfunction $f(x, y, t )$ on the physical space
may be constructed as
\begin{equation}
    f(x, y, t) = \phi(x, y, t) + \phi( -x, -y, t)\,.
\end{equation}

In order to find the functions on Gott spacetime
which is generated by two relatively
moving point masses, this condition must be expressed in a frame in which the
point mass is moving.
The Lorentz transform to the moving frame it is most conveniently done in
null coordinates
\begin{equation}
x^{\pm} \equiv x \pm t\,;
\end{equation}
then, in boosted null coordinates,
\begin{eqnarray}
x^{\prime\pm} &=& e^{\pm\alpha} x^{\pm}\,, \nonumber\\
y^{\prime} &=& y - Y_0\,.
\end{eqnarray}
The mass point which was at $x = 0 = y$ in the original coordinates
is at $x' - t' \tanh\alpha = 0 = y' + Y_0$ in the new coordinates,
corresponding to a mass point moving with velocity $\tanh\alpha$ in the new
coordinates.
Since the metric was Minkowskian in the original coordinates, it is
still the Minkowski metric but with the boosted identification
\begin{equation}
    x^{\pm} \doteq  - e^{\pm 2\alpha} x^{\mp}\,, \qquad
    \mbox{at}\qquad y = - Y_0\,,
    \label{lower-identification}
\end{equation}
where the primes on the coordinates have been dropped.
The condition that a function on the covering space defines a function on
the physical space is
\FL
\begin{equation}
f( x^+, x^-, y ) = f( - e^{2\alpha} x^-, - e^{-2\alpha} x^+, - y - 2 Y_0  )\,,
\label{lowercondition}
\end{equation}
where $f$ is now regarded as a function of the null coordinates $x^{\pm}$
rather than as a function of $x$ and $t$.

If a second mass point
moving in the opposite direction is added at $y = Y_0$, there
is then the further identification
\begin{equation}
    x^{\pm} \doteq  - e^{\mp 2\alpha} x^{\mp} \,, \qquad
    \mbox{at}\qquad y = Y_0\,.
    \label{upper-identification}
\end{equation}
The resultant space shown in Fig.\ \ref{fig:2mass} is restricted to the
region, $ -Y_0 < y < Y_0$.
For a total deficit angle less than $2\pi$,
the space is asymptotically a cone.
In this case, the cone has zero opening angle, {\it i.e.\/} it is
asymptotically one end of a cylinder for point masses relatively at rest.
Because of the relative motion of the point masses, the cylinder must be
regarded as two cylinder halves joined with a relative boost.

The continuity conditions for functions on this space are those of the
original string, Eq.\ (\ref{lowercondition}), as well as
\FL
\begin{equation}
f( x^+, x^-, y ) =
f( - e^{ - 2 \alpha} x^-, - e^{ 2 \alpha} x^+, - y + 2 Y_0  )\,;
\label{uppercondition}
\end{equation}
these may be combined to yield the condition
\begin{equation}
f( x^+, x^-, y ) =
f( e^{- 4\alpha} x^+, e^{4\alpha} x^-, y + 4 Y_0  )\,,
\label{wrapcondition}
\end{equation}
for a function on the covering space to define a function on the physical
space.
This condition may be combined with either of the two previous conditions to
form a sufficient set of conditions for a function on the covering space to
define a function on the physical space.
When  the rapidity $\alpha$ vanishes, the masses are stationary,
and the functions are periodic in $y$ and even under
inversion in the location of each of the masses.
In the general case, the functions are even under boosted inversion
in the location of each of the masses which implies that they are periodic
under
simultaneous translations in $y$ by $4Y_0$ and in rapidity by $-4\alpha$.

Just as with the single mass, an arbitrary function $\phi(x^+, x^-, y )$
on the covering space may be used to generate a function on the physical
space which explicitly satisfies the continuity conditions,
Eqs.\ (\ref{lowercondition}, \ref{uppercondition}, \ref{wrapcondition}),
\widetext
\ifpreprintsty
\begin{eqnarray}
f(x^+, x^-, y ) & = & \sum_{n= -\infty}^{\infty}
    \left[
    \phi(e^{- 4 n \alpha}x^+, e^{4 n \alpha} x^-, y + 4 n Y_0) \right.
     \nonumber\\
 & + & \left.
    \phi( - e^{ - (4n + 2 ) \alpha} x^-,
    - e^{(4n + 2)\alpha} x^+, -y + (4n +2) Y_0 ) \right]\!,
\end{eqnarray}
or
\begin{equation}
f(x^+, x^-, y ) = \sum_{n = -\infty}^{\infty} \phi( x^+_n, x^-_n, y_n )
\,,
\label{generalfunction}
\end{equation}
\else
\FL
\begin{eqnarray}
f(x^+, x^-, y ) & = & \sum_{n= -\infty}^{\infty}
    \left[
    \phi(e^{- 4 n \alpha}x^+, e^{4 n \alpha} x^-, y + 4 n Y_0) +
    \phi( - e^{ - (4n + 2 ) \alpha} x^-,
    - e^{(4n + 2)\alpha} x^+, -y + (4n +2) Y_0 ) \right]
    \nonumber
    \\
 & = & \sum_{n = -\infty}^{\infty} \phi( x^+_n, x^-_n, y_n ) \,,
\label{generalfunction}
\end{eqnarray}
\fi
\narrowtext
where
\begin{equation}
\begin{array}{rcl}
x^{\pm}_n & = & \left\{
		    \begin{array}{rl}
		    e^{ \mp 2 n \alpha} x^{\pm}\,, &\mbox{\quad$n$ even,} \\
		    - e^{ \mp 2 n \alpha} x^{\mp}\,, &\mbox{\quad$n$ odd,}
		    \end{array}
		\right. \\
	\noalign{\vspace{0.5ex}}
	y_n & = & \left\{
		    \begin{array}{rl}
		    y + 2 n Y_0\,, &\mbox{\quad$n$ even,} \\
		    - y + 2 n Y_0\,, &\mbox{\quad$n$ odd.}
		    \end{array}
		\right.
		\end{array}
\label{eq:images}
\end{equation}

Two cases of particular interest are the point source and the Green's function
which it produces.
On the covering space a single source at the spacetime
point $x'$ is described by
\FL
\begin{eqnarray}
\delta^{(3)}(x - x') & = &
    \delta(t - t')\delta(x - x')\delta(y - y')\nonumber \\
    & = & 2 \delta(x^+ -x'^+) \delta(x^- -x'^-)\delta(y - y')\,.
\end{eqnarray}
The point source on the physical space is thus given by
\ifpreprintsty
\begin{equation}
\delta_p^{(3)}( x - x') =
\sum_{n= -\infty}^{\infty}
    2 \delta( x^+ - x'^+_n)\delta(x^- - x'^-_n)\delta(y - y'_n)\,.
    \label{eq:delta}
\end{equation}
\else
\begin{eqnarray}
\lefteqn{\delta_p^{(3)}( x - x') = } \nonumber \\
& &
\sum_{n= -\infty}^{\infty}
    2 \delta( x^+ - x'^+_n)\delta(x^- - x'^-_n)\delta(y - y'_n)\,,
    \label{eq:delta}
\end{eqnarray}
\fi
and the source in the physical spacetime
is represented as a sum of image sources in the covering spacetime.
Note that this expression is symmetric in $x$ and $x'$ so that either may be
taken to be the independent variable, with the other being the location of the
source point.
As a result, this delta function
defined on the covering space is a good function on the physical space when
regarded as either a function of $x$ or of $x'$.
In $3+1$ dimensions, the delta function has an additional overall factor of
$\delta(z - z')$.

The Green's function equation
\begin{equation}
(-\partial^2 + m^2)G(x,x') = \delta(x - x')\,,
\end{equation}
has the solution
\FL
\begin{equation}
G_0( x, x ') = \left\{
	\begin{array}{ll}
	\displaystyle i \exp({-m s(x,x')})
	\over
	\displaystyle 4\pi s(x,x') & \mbox{\qquad $2+1$ dim}\\
	\noalign{\vspace{0.3ex}}
	\displaystyle i m K_1(m s(x,x') ) \over
	\displaystyle 2 \pi^2 s(x,x') &
	    \mbox{\qquad $3+1$ dim}
	\end{array}
    \right.
\end{equation}
in the covering space, where
$s(x,x') = \sqrt{ (x-x')^2}$, and the choice of analytic continuation
into the region where $s < 0$
determines which Green's function is obtained.
The solution in the physical space then has the image form
\begin{equation}
G(x,x') = \sum_{n = -\infty}^{\infty} G_0(x, x'_n)\,.
\label{eq:green-sum}
\end{equation}
The choice of analytic continuation is critical; it will be discussed at
length in Sec.\ \ref{Green}.

\section{Acausality}
\label{acausal}

In the preceding section, a global set of coordinates for the
spacetime with two moving point masses were found:
The metric is the flat metric over
the region $ -Y_0 < y < Y_0$ with the identifications given by
Eqs.\ (\ref{lower-identification}), (\ref{upper-identification}).
Since the causal properties of such spacetimes have been discussed extensively
\cite{Gott,Ori,Cutler}, a brief discussion will suffice here.

The conclusion is that this spacetime consists of three regions:
A future region, a past region, and an acausal region.
The past and future regions are causal in that
no timelike or null curves intersect
themselves in those regions, and they are respectively bounded in the future
and the past by the Cauchy horizon which separates them from the acausal
region.

The spacetime may be embedded in the covering spacetime,
the causal properties of which are trivial.
The source of a timelike or null curve has an infinite number of images in the
covering space, Eq.\ (\ref{eq:images}), and a given spacetime point is within
the future light cone of the source point in the physical spacetime if it is
within the future light cone of any one of the images in the covering
spacetime.
A given point in the physical space may be identified with the $n=0$ image in
the covering space.
The point can be connected to itself by a non-spacelike curve in the
physical space if and only if the image of the curve in the
covering space is a non-spacelike curve.
If the path in the physical space wraps $n'$ times around the world
line(s) of the point masses, its image in the covering space goes from
the $n=n'$ image of the source point to the identified, $n=0$, image
point.
\ifprinter
\ifpreprintsty
\begin {figure}[htbp]%			Figure 3
    \vbox to 2.3in{
	\vss
	%\vskip -0.75in
	\leavevmode {\hfill \epsfbox {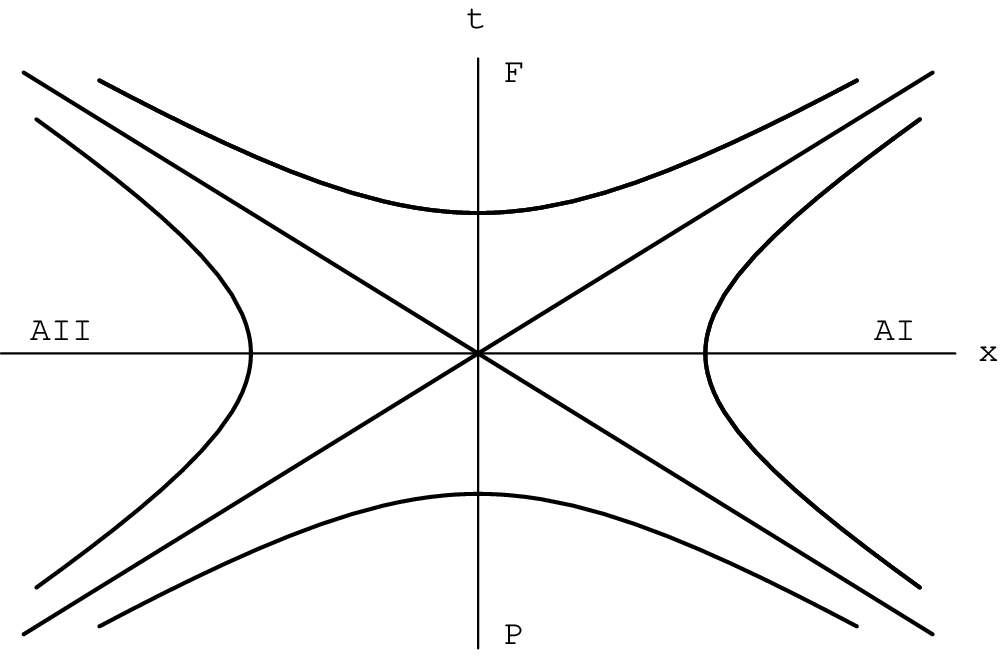} \hfill}
	%\vskip -0.75in
	\vss
	}
    \caption
	{%
	\advance\baselineskip by -8pt
	The hyperbolae show the surfaces on which images of a point in the
	physical region lie.
	The $y$ axis is perpendicular to the graph, and the physical space
	consists of the region $-Y_0 < y < Y_0$.
	}%
    \label {fig:hyperbolae}
\end {figure}
\fi
\fi

Consider the past region P shown in Fig.\ \ref{fig:hyperbolae}.
A source point in P has $x^+ < 0 < x^-$, and its images at
$(x^{\pm}_n, y_n)$
all lie on the same hyperbolic surface but displaced in $y$.
The hyperbolic surface
$x^+ x^- = \mbox{const} < 0$, $x^- > 0$, and $-\infty < y < \infty$
is a spacelike surface so that all the images have spacelike
separations from each other in the covering space.
Thus, no future directed non-spacelike curve can connect the source to any of
its images in the covering space, and, therefore, no future directed
non-spacelike curve can connect the source point to itself in the physical
space.
A similar argument holds for the future region F defined by $x^+ > 0 >
x^-$. Hence both regions are causal, {\it i.e.\/} no non-spacelike curve
can intersect itself in either region.

The argument does not hold for the acausal region A shown as two separate
regions AI and AII in Fig.\ \ref{fig:hyperbolae}.
(Note that although the two regions appear to be
separated, they are not since they
are connected at $y = \pm Y_0$.)\ \ Since these regions are defined by,
respectively, $x^\pm > 0$, and $x^\pm < 0$,
the images given by Eq.\ (\ref{eq:images}) appear alternately in the
left and right branches on the hyperbolae shown.
In addition, the images are displaced in $y$ perpendicular to the diagram.
The two surfaces in the covering space on which the images lie are each
timelike surfaces which are asymptotically null at infinity, and
each surface is everywhere spacelike relative to the other surface.
As a result, no non-spacelike curve can connect a point to an image on the
reflected hyperbola, but, since each surface is timelike, the images on the
same hyperbola can in general be connected to each other by a non-spacelike
curve.
That is, non-spacelike curves which start out in A and wrap around both masses
some number of times may intersect themselves.
Because successive images are displaced in $y$, there is a spacelike component
of the separation in the $y$ direction.
As a result, the projection of the
lightcones into the $x$-$t$ plane is narrower than $45^{\circ}$, and some, but
not all, images can be connected by a non-spacelike curve.

Points in the acausal region can actually be connected to themselves
by non-spacelike
goedesics.
To see this, note that the image of the geodesic in the physical space is the
usual straight line in the covering space.
The tangent vector of the curve is $\vec \partial / \partial \tau = ( k^+,
k^-, \pm 1)$, where the affine parameter $\tau$ is normalized so that
$dy/d\tau = \pm 1$ in the covering space.
The geodesic is a future directed non-spacelike curve if $k^+ > 0 > k^-$, and
$ - k^+ k^- \geq 1$.
Thus $k^{\pm} = \pm e^{\pm\eta} v$, where $v \ge 1$.
The curve starts at the image point given by (\ref{eq:images}),
$(e^{-4n\alpha}x^+, e^{4n\alpha}x^-, y + 4 n Y_0)$, and then, in the covering
space, is given by
\begin{equation}
    \begin{array}{rcl}
	x^+(\tau) & = & e^{-4 n \alpha} x^+ + v e^{\eta} \tau \,, \\
	x^-(\tau) & = & e^{4 n \alpha} x^- - v e^{-\eta} \tau \,, \\
	y( \tau)  & = & y + 4 n Y_0 \pm \tau\,.
    \end{array}
\label{eq:geodesic}
\end{equation}
The geodesic reaches $(x^{\pm},y)$ provided there is a value of
$\tau$ such that
$x^+(\tau) = x^+$, $x^-(\tau) = x^-$, and $y(\tau) = y$.
A solution exists only if the plus (minus) sign is chosen in
(\ref{eq:geodesic}), and $n$ and $x^{\pm}$ are all negative (positive).
In the case where the initial point is in the left segment AII, let $x^{\pm}
= -X e^{\pm \beta}$, $ n = -N$, and the solution is
\begin{equation}
\begin{array}{rcl}
\tau & = & 4 N Y_0 \,,\\
     v  & = & { X \sinh 2 N \alpha / ( 2 N Y_0 ) } \,,\\
e^{\eta}& = & e^{\beta + 2 N \alpha}\,.
\end{array}
\label{eq:geodesic-soln}
\end{equation}
The number $N$ is just the winding number of the path which connects the point
$(x^{\pm},y)$ to itself.
Typical curves for winding numbers 1 and 2 are shown in the physical space
in Fig.\ \ref{fig:ctcs}.
A future directed non-spacelike curve with winding number $N$ connects the
point $x$ with itself in the physical space provided that
the curve is non-spacelike.
This is true if $v = X\sinh 2N\alpha/2NY_0 \geq 1$.
For large enough winding number every point with $X > 0$ can be
connected to itself.
Thus, the acausal region consists of all those points with
$x^+ x^- > 0$.
The Cauchy horizons are, by definition, the boundary of the acausal region;
these are the null surfaces $x^+ = 0$ and $x^- = 0$.
The Cauchy horizons are not part of the acausal region because the images of a
source point on the Cauchy horizon are all spacelike with respect to the
source point, hence no non-spacelike curve can connect the image to the source
point.

\ifprinter
\ifpreprintsty
\begin {figure}[htbp]%			Figure 4
    \vbox to 2.3in{
	\vss
	%\vskip -0.75in
	\leavevmode {\hfill \epsfbox {ctcs.eps} \hfill}
	%\vskip -0.75in
	\vss
	}
    \caption
	{%
	\advance\baselineskip by -8pt
	The left and right figures respectively show closed future directed
	timelike curves of winding number 1 and 2.
	The identified points are labeled in increasing
	temporal order from A to B (D) along the
	respective curves.
	}%
    \label {fig:ctcs}
\end {figure}
\fi
\fi

The surfaces for which $v = 1$ ($X = 2NY_0/\sinh2N\alpha$) are the surfaces
consisting of points at which null geo\-desics intersect themselves;
Kim and Thorne \cite{Thorne:vacuum} refer to them as polarized hypersurfaces.
They will be discussed further in Sec.\ \ref{stress}, where  it will be shown
that they are the loci of weak singularities of the stress-energy.

The surfaces $x^+x^- = \mbox{const} < 0$ are spacelike surfaces restricted to
the causal region $F$ ($P$) if $x^+ > 0$ ($x^+ < 0$); however, they are not
Cauchy surfaces because there are future directed non-spacelike curves which
do not intersect the surfaces. (For example, a null curve coming in along
the Cauchy horizon does not intersect a surface in the past region.)
It is true that every timelike geodesic intersects every such surface in both
the past and the future regions and, as will be discussed in Sec.\ \ref{eigen},
arbitrary solutions to the wave equation may be specified in terms of data on
these surfaces.

%\input{eigen}
%\ifpreprintsty\pagebreak\fi
\section{Eigenfunctions}
\label{eigen}

For various purposes it is convenient to have an explicit set of complete
orthonormal functions on the physical space; it is even more convenient if
they are eigenfunctions of the wave operator.
In the case of the Gott space considered here, it is easy to construct
eigenfunctions on the covering space.
To construct a complete set of orthonormal eigenfunctions of the wave
operator on the physical space is not much more difficult.
The main problem is
showing that they are, in fact, a complete set.

On the covering space the eigenfunction $\psi$ of the wave operator satisfies
the equation
\begin{equation}
\partial^2 \psi(x^+, x^-, y ) = w \psi(x^+, x^-, y ) \,,
\label{eq:waveeq}
\end{equation}
where $w$ is the eigenvalue of the wave operator.
The general eigenfunction is given by
\FL
\begin{eqnarray}
\psi(x^+, x^-, y )
& = \frac{1}{2} &
\int\,dk^+\,dk^-\,dk_y e^{i(k^+x^- + k^-x^+ + k_y y)}
\nonumber \\
& &
\delta( k^+k^- + k_y^2 + w) f(k^+, k^-, k_y)\,.
\end{eqnarray}
A special case is
\FL
\begin{eqnarray}
\lefteqn{\psi_{w,k_y}(x^+, x^-, y )
 = \frac{1}{2} \int\,dk^+\,dk^- e^{i(k^+x^- + k^-x^+)/2}}
 \hspace{ 5em}
\nonumber \\
 & &  \delta( k^+k^- + k_y^2 + w)
    e^{ik_y y} f_{w,k_y}(k^+/ k^-) \,,
\end{eqnarray}
where $f$ now depends only on the ratio $k^+/k^-$.
This function must be invariant under the transformation,
(\ref{wrapcondition}), $x^{\pm} \rightarrow
e^{\mp4\alpha}x^{\pm}$ and $y \rightarrow y + 4 Y_0$, or
\FL
\begin{equation}
 f_{w,k_y}(k^+/ k^-) =
e^{i k_y  4 Y_0} f_{w,k_y}(e^{- 8\alpha}k^+/ k^-)\,,
\end{equation}
which implies that a special solution with periodicity $n$ and homogeneous in
$k^+/k^-$ with power $i\eta$ is
\begin{equation}
f_{w, k_y}( k^+/k^-) =
e^{i{y\over Y_0}({n\pi\over 2} + \alpha\eta)} (k^+/k^-)^{i\eta/2}\,.
\end{equation}
The general solution is then a superposition of the special solutions
\FL
\begin{eqnarray}
\lefteqn{\psi_{w',\eta,n}(x^+, x^-, y ) =
 {1\over  4\pi}\int\,dk^+dk^- e^{i(k^+ x^- + k^- x^+)/2} } \hspace{5em}
\nonumber \\
 & &
 \delta( k^+k^- + w')
e^{i{y\over Y_0}({n\pi\over 2} + \alpha\eta)} (k^+/k^-)^{i\eta/2} \!,
\end{eqnarray}
where $w' = w + ( n \pi/2 + \alpha\eta)^2/Y_0^2$ and $k_y = (n\pi/2 + \alpha
\eta)/Y_0$.
The function still does not satisfy the reflection conditions,
(\ref{lowercondition},\ref{uppercondition}), around the world
lines of the
two masses separately,
and the ranges of the $k^{\pm}$ integrations are not yet
determined.
Under the reflections
$x^{\pm} \rightarrow  - e^{\mp2\alpha}x^{\mp}$ and
$y       \rightarrow  - y + 2 Y_0$,
the corresponding change of variables
\begin{equation}
k^{\pm} \rightarrow  - e^{\mp2\alpha}k^{\mp} \,,
\label{eq:k}
\end{equation}
which, along with $n \rightarrow  - n$ and $\eta \rightarrow  - \eta$,
yields,
modulo the question of the phases in the factor $(k^+/k^-)^{i\eta/2}$,
\FL
\begin{equation}
\psi_{w',\eta,n}(x^+, x^-, y ) \rightarrow
	    ( -1 )^n \psi_{w', -\eta, -n}(x^+, x^-, y)\,.
\end{equation}

If $w' > 0$, the two dimensional vector $(k^+,k^-)$ is timelike, and the
transformation (\ref{eq:k}) preserves the signs of $k^{\pm}$.
The positive frequency solution may then be taken to be
\widetext
\begin{eqnarray}
\psi_{w',\eta,n}^{(+)}(x^+, x^-, y) & = &
{1\over (2\pi)^{3/2}}\frac{1}{2}
\int^{\infty}_0\,dk^+ \int^{\infty}_{-\infty}\,dk^-
\delta( k^+ k^- + w') e^{i(k^+x^- + k^-x^+)/2} \nonumber \\
& &  \qquad \left[
{e^{i{y\over Y_0}({n\pi\over2} + \alpha\eta)} \over \sqrt{2 Y_0}i^n}
\left(k^+\over - k^-\right)^{i\eta/2} +
{e^{- i{y\over Y_0}({n\pi\over2} + \alpha\eta)} \over \sqrt{2 Y_0}i^{-n}}
\left(k^+\over - k^-\right)^{-i\eta/2} \right] \!,
\label{eq:psi+}
\end{eqnarray}
where,
due to the explicit minus sign  in the $k^+/(-k^-)$ factor, there are no phase
ambiguities.
The expression for negative $\eta$ and $ n \rightarrow - n$ is
the same as Eq.\ (\ref{eq:psi+}), hence $\eta$ may be taken to be positive.
The overall normalization is shown to be correct in
Appendix \ref{app:orthonormal}, where the orthonormality and completeness of
the functions are shown.

The complex conjugate of $\psi^{(+)}$ is a negative frequency solution which
is independent of the positive frequency solution; it may be written in the
form
\goodbreak
\ifpreprintsty
\begin{eqnarray}
\lefteqn{\psi_{w',\eta,n}^{(-)}(x^+, x^-, y)  =
\left(\psi_{w',\eta,n}^{(+)}(x^+, x^-, y)\right)^{\ast} = } \qquad
\nonumber \\
 & & \qquad{1\over (2\pi)^{3/2}} \frac{1}{2}\int^{\infty}_0\,dk^+
\int^{\infty}_{-\infty}\,dk^-
\delta( k^+ k^- + w') e^{-i(k^+x^- + k^-x^+)/2} \nonumber \\
 & & \qquad \qquad\left[
{e^{i{y\over Y_0}({n\pi\over2} + \alpha\eta)} \over \sqrt{2 Y_0}i^n}
\left(k^+\over - k^-\right)^{i\eta/2} +
{e^{- i{y\over Y_0}({n\pi\over2} + \alpha\eta)} \over \sqrt{2 Y_0}i^{-n}}
\left(k^+\over - k^-\right)^{-i\eta/2}
\right]\!.
\label{eq:psi-}
\end{eqnarray}
\else
\begin{eqnarray}
\lefteqn{\psi_{w',\eta,n}^{(-)}(x^+, x^-, y)  =
\left(\psi_{w',\eta,n}^{(+)}(x^+, x^-, y)\right)^{\ast} =
{1\over (2\pi)^{3/2}}
\frac{1}{2}\int^{\infty}_0\,dk^+ \int^{\infty}_{-\infty}\,dk^-
\delta( k^+ k^- + w')} \hspace {5em} \nonumber \\
& & e^{-i(k^+x^- + k^-x^+)/2}
\left[
{e^{i{y\over Y_0}({n\pi\over2} + \alpha\eta)} \over \sqrt{2 Y_0}i^n}
\left(k^+\over - k^-\right)^{i\eta/2} +
{e^{- i{y\over Y_0}({n\pi\over2} + \alpha\eta)} \over \sqrt{2 Y_0}i^{-n}}
\left(k^+\over - k^-\right)^{-i\eta/2}
\right]\!.
\label{eq:psi-}
\end{eqnarray}
\fi

For $w' < 0$ the momentum $(k^+, k^-)$ is spacelike, and the
transformation, Eq.\ (\ref{eq:k}), changes the sign of $k$; instead of
Eq.\ (\ref{eq:k}), the change of variables
\begin{equation}
k^{\pm} \rightarrow  e^{\mp2\alpha}k^{\mp} \,,
\label{eq:k-}
\end{equation}
is used, and the solution to the wave equation on the physical space,
Eq.\ (\ref{eq:waveeq}), is
\ifpreprintsty
\begin{eqnarray}
\psi_{ w',\eta,n}^{s}(x^+, x^-, y) & = &
 {1\over (2\pi)^{3/2}} \frac{1}{2}\int^{\infty}_0\,dk^+
\int^{\infty}_{-\infty}\,dk^-
\delta( k^+ k^- + w') \nonumber \\
 & &  \hspace{- 5em} \left[
{e^{i(k^+x^- + k^-x^+)/2}
e^{i{y\over Y_0}({n\pi\over2} + \alpha\eta)} \over \sqrt{2 Y_0}i^n}
\left(k^+\over k^-\right)^{i\eta/2} \right. \qquad
\label{eq:psi-s} \\
 & + & \left. {e^{-i(k^+x^- + k^-x^+)/2}
 e^{- i{y\over Y_0}({n\pi\over2} + \alpha\eta)} \over \sqrt{2 Y_0}i^{-n}}
\left(k^+\over k^-\right)^{-i\eta/2} \right] \!, \nonumber
\end{eqnarray}
\else
\begin{eqnarray}
\psi_{ w',\eta,n}^{s}(x^+, x^-, y) & = &
 {1\over (2\pi)^{3/2}} \frac{1}{2}\int^{\infty}_0\,dk^+
\int^{\infty}_{-\infty}\,dk^-
\delta( k^+ k^- + w')  \nonumber \\
 & &  \hspace{- 5em} \left[
{e^{i(k^+x^- + k^-x^+)/2}
e^{i{y\over Y_0}({n\pi\over2} + \alpha\eta)} \over \sqrt{2 Y_0}i^n}
\left(k^+\over k^-\right)^{i\eta/2} +
{e^{-i(k^+x^- + k^-x^+)/2}
 e^{- i{y\over Y_0}({n\pi\over2} + \alpha\eta)} \over \sqrt{2 Y_0}i^{-n}}
\left(k^+\over k^-\right)^{-i\eta/2} \right] \!,
\label{eq:psi-s}
\end{eqnarray}
\fi
where $w' < 0$.
This solution is real, and changing the sign of $\eta$ does not yield an
equivalent solution, hence $-\infty < \eta < \infty$.

As is shown in Appendix \ref{app:orthonormal} these functions
form a complete orthonormal set,
\begin{equation}
\int \,dx \, \psi^{a'}_{w',\eta',n'}(x)^{\ast}
\psi^{a}_{w,\eta,n}(x) =
\delta_{a',a}\delta( w' - w)\delta(\eta' - \eta)\delta_{n',n}\,,
\label{eq:ortho}
\end{equation}
for integration over the physical space, and
\begin{eqnarray}
\lefteqn{2\delta( x'^+ - x^+)\delta( x'^- - x^-)\delta(y' - y) = }
\nonumber \\
&\qquad &
\int_0^{\infty}\,dw' \int_0^{\infty}\,d\eta \sum_{n = -\infty}^{\infty}
\left(
\psi^{(+)}_{w',\eta,n}(x')^{\ast}
\psi^{(+)}_{w',\eta,n}(x) +
\psi^{(-)}_{w',\eta,n}(x')^{\ast}
\psi^{(-)}_{w',\eta,n}(x) \right) \nonumber \\
&\qquad &
\null + \int^0_{- \infty}\,dw' \int_{-\infty}^{\infty}\,d\eta
\sum_{n = -\infty}^{\infty}
\psi^{s}_{w',\eta,n}(x')^{\ast}
\psi^{s}_{w',\eta,n}(x) \,,
\label{eq:ortho1}
\end{eqnarray}
for $x'$ and $x$ both in the physical space.

As a corollary to these results, an arbitrary function, $\phi$,
on the physical space may be written as
\begin{eqnarray}
\phi(x) & = & \sum_{n = -\infty}^{\infty} \int_0^{\infty}\,dw'
\int_0^{\infty}\,d\eta
\left(\psi^{(+)}_{w',\eta,n}(x) f_n^{(+)}(w', \eta)
+ \psi^{(-)}_{w',\eta,n}(x) f_n^{(-)}(w', \eta)\right)
\nonumber \\
 &+&\sum_{n = -\infty}^{\infty} \int^0_{-\infty}\,dw'
\int_{-\infty}^{\infty}\,d\eta
\psi^{s}_{w',\eta,n}(x) f_n^{s}(w', \eta) \,.
\label{eq:expand}
\end{eqnarray}

The solutions to the wave equation, Eq.\ (\ref{eq:waveeq}), for a given value
of
$w = m^2$ are, of course, not complete on the spacetime.
However, they are complete on a given spacelike surface in Minkowski
spacetime.
That is, using positive and negative frequency solutions,
a solution with an arbitrary initial value and
an arbitrary initial time derivative on the surface can be constructed.
Alternatively, using positive frequency solutions,
a positive frequency solution with an
arbitrary initial value (or an arbitrary initial time derivative) can be
constructed.
In the Gott space, the same is true with some qualifications.
An arbitrary solution can be constructed
for an arbitrary spacelike surface solely within either the future region
or the past region.
That is, an arbitrary value of either the function or its normal derivative may
be specified.
If both positive and negative solutions are used, both the value of the
function and its normal derivative may be specified.

The positive frequency solution for the wave equation is
\ifpreprintsty%\pagebreak
\begin{eqnarray}
\lefteqn{\phi_{\eta,n}^{(+)}(x^+, x^-, y) \equiv \sqrt{2\pi}
\psi^{(+)}_{w, \eta, n}(x^+, x^-, y) =
 {1\over (2\pi)}\frac{1}{2} \int^{\infty}_0\,dk^+ \int^{\infty}_{-\infty}\,dk^-
} \hspace{ 3em}
\nonumber \\
& &
\delta( k^+ k^- + m^2 + (n\pi/2 + \alpha\eta)^2/Y_0^2 )
e^{i(k^+x^- + k^-x^+)/2}
\nonumber \\
 & &
 \left[
{e^{i{y\over Y_0}({n\pi\over2} + \alpha\eta)} \over \sqrt{2 Y_0}i^n}
\left(k^+\over - k^-\right)^{i\eta/2} +
 {e^{- i{y\over Y_0}({n\pi\over2} + \alpha\eta)} \over \sqrt{2 Y_0}i^{-n}}
\left(k^+\over - k^-\right)^{-i\eta/2} \right] \!,
\label{eq:phi+}
\end{eqnarray}
\else
\FL
\begin{eqnarray}
\lefteqn{\phi_{\eta,n}^{(+)}(x^+, x^-, y) \equiv \sqrt{2\pi}
\psi^{(+)}_{w, \eta, n}(x^+, x^-, y) =
 {1\over (2\pi)}\frac{1}{2} \int^{\infty}_0\,dk^+ \int^{\infty}_{-\infty}\,dk^-
\delta( k^+ k^- + m^2 + (n\pi/2 + \alpha\eta)^2/Y_0^2 )
} \hspace{ 5em}
\nonumber \\
& &
e^{i(k^+x^- + k^-x^+)/2}
 \left[
{e^{i{y\over Y_0}({n\pi\over2} + \alpha\eta)} \over \sqrt{2 Y_0}i^n}
\left(k^+\over - k^-\right)^{i\eta/2} +
 {e^{- i{y\over Y_0}({n\pi\over2} + \alpha\eta)} \over \sqrt{2 Y_0}i^{-n}}
\left(k^+\over - k^-\right)^{-i\eta/2} \right] \!,
\label{eq:phi+}
\end{eqnarray}
\fi
\narrowtext\noindent
where $w = m^2 + (n\pi/2 + \alpha\eta)^2/(4Y_0)^2$,
and the conservation condition for solutions to the wave equation reads
\begin{equation}
\partial_{\nu}\left[
\phi'(x)
\mathrel{\mathop{\partial^{\nu}}\limits^{\leftrightarrow}}
\phi(x)\right] = 0\,.
\end{equation}
Then the integral over a surface
\begin{equation}
(\phi',\phi) = \int\,d\sigma_{\nu}
\phi'(x)
{1\over i}\mathrel{\mathop{\partial^{\nu}}\limits^{\leftrightarrow}}
\phi(x) \,,
\label{eq:surface-integral}
\end{equation}
is constant, provided that the functions $(\phi',\phi)$ drop off fast enough
at infinity.
In Appendix\ \ref{app:orthonormal}, this integral, in the case
of the basis functions (\ref{eq:phi+}) and spacelike surfaces restricted
to either the future or the past, is shown to be just the orthonormality
relation.

The integral over a surface is most easily done by changing into coordinates
appropriate to the surface, $x \rightarrow (\tau, \xi)$, where $\tau$ is the
coordinate labeling the surface, and $\xi$ are the coordinates in the surface.
The normal to the surface is $ \vec n = \vec \partial/\partial\tau$, and the
integral becomes
\begin{equation}
(\phi',\phi) = \int\,d\xi n_{\nu}\sqrt{-g}
\phi'(x)
{1\over i}\mathrel{\mathop{\partial^{\nu}}\limits^{\leftrightarrow}}
\phi(x) \,,
\end{equation}
where $-g$ is the absolute value of the determinant of the metric.
It is shown in Appendix \ref{app:orthonormal} that, for an arbitrary $x^+x^- =
\mbox{constant}$ surface in either the past or the future region,
\FL
\begin{equation}
\int\,d\sigma_{\nu}
\phi^{(+)}_{\eta,n}(x)^{\ast}
{1\over i}\mathrel{\mathop{\partial^{\nu}}\limits^{\leftrightarrow}}
\phi'^{(+)}_{\eta',n'}(x) = \delta(\eta - \eta')\delta_{n,n'}\,.
\end{equation}
Thus an arbitrary positive frequency solution to the wave equation,
\begin{equation}
\phi(x) = \int_0^{\infty}\,d\eta \sum_{n = -\infty}^{\infty}
\phi^{(+)}_{\eta,n}(x) \phi(\eta,n) \,,
\end{equation}
has the same (positive definite) norm in both the past and in
the future regions,
\begin{equation}
\left(\phi',\phi\right) =
 \int_0^{\infty}\,d\eta \sum_{n = -\infty}^{\infty}
\phi'(\eta,n)^{\ast}\phi(\eta,n)\,.
\end{equation}
That is, every particle which starts out in the past region eventually ends up
in the future region, and every particle which ends up in the future region
started out in the past region.

\section{Green's Function}
\label{Green}

The formulation of a quantum field is well understood in Minkowski space.
There are several alternate formulations: one may use the Wightman functions,
the time-ordered product, the Green's function, or a functional integral to
calculate the matrix elements of the field, its spectrum, and any scattering
which may occur.

In summary, the basic quantity is the Wightman function
\begin{eqnarray}
\Delta^{(+)}(x,x')
& = &
\left\langle 0+ \left| \phi(x) \phi(x') \right| 0- \right\rangle
\nonumber \\
& = & \int {d^3p \over 2 p^0 } { e^{ip(x -x')} \over (2\pi)^3 }\,,
\label{eq:Wightman:M}
\end{eqnarray}
where the initial and final states are respectively the initial and final
vacua.
In the case of the free field in Minkowski space, these vacua are the same.
The frequency $p^0 = \sqrt{\vec p^2 + m^2}$.
Since the vacuum is the lowest energy state, the matrix element has positive
(negative) frequency with respect to $t\ (t')$.
The time-ordered product
\begin{eqnarray}
\Delta_F(x,x')
 & = & i \left\langle 0+ \left|
( \phi(x) \phi(x') )_+
\right| 0- \right\rangle
\nonumber \\
& = &
i\theta( t - t')
\left\langle 0+ \left| \phi(x) \phi(x') \right| 0- \right\rangle
\label{eq:timeordered:M}
\\
& + &
i\theta( t' - t)
\left\langle 0+ \left| \phi(x') \phi(x) \right| 0- \right\rangle \!,
\nonumber
\end{eqnarray}
both yields the Wightman function and can be constructed from it.
It satisfies the Green's function equation
\begin{equation}
\left(-\partial^2 + m^2\right) \Delta_F(x,x') = \delta( x - x') \,,
\label{eq:green's equation}
\end{equation}
with the boundary condition that it be of positive (negative) frequency for
$t > t'$ ($t < t'$).
This equation and boundary condition uniquely determine the Green's function
to be
\begin{equation}
\Delta_F(x,x') = \int\,{d^4k\over (2\pi)^4}
{e^{ikx}\over k^2 + m^2 -i\epsilon }\,,
\label{eq:green's function:M}
\end{equation}
where the $- i \epsilon$ enforces the positive frequency boundary conditions.
This expression in turn may be represented as the functional integral over all
field configurations on $M^4$,
\begin{equation}
\Delta_F(x,x') =  i \int\,\left[d\phi\right] e^{iW[\phi]} \phi(x)\phi(x')\,,
\label{eq:functional integral}
\end{equation}
where
\begin{equation}
W[\phi] = - \frac{1}{2}\int\,d^4x
\left[
(\partial\phi)^2
+(m^2 - i \epsilon)\phi^2
\right]\!.
\label{eq:action}
\end{equation}
The $- i \epsilon$ in the integrand is needed to allow an analytic
continuation to complex times which enforces convergence of the functional
integral and also yields the correct boundary conditions on $\Delta_F$.
The validity of this formulation may be established either
1) by explicitly doing
the functional integral or
2 ) by summing over all possible field configurations at each time
(Cauchy surface) using the Hamiltonian to propagate the fields from
one time to the next.

In Gott space the above line of argument does not work.
The Wightman function (\ref{eq:Wightman:M}) must be a solution to the
homogeneous equation.
If the state $\langle 0+ |$ ($| 0- \rangle$) is assumed to be the
lowest energy state in the future (past), then the Wightman function
must be a positive frequency function; hence it must be expressible as a
superposition of the functions
$\phi^{(+)}_{\eta, n}(x)$ ($\phi^{(-)}_{\eta, n}(x')$).
Because the mass points are moving, the space is not static, and there
is no conserved energy; nonetheless, for a free field there is a conserved
particle number, and the zero particle state has zero energy which is less
than that of any other state.
Since there is particle conservation, one can require
that the Wightman function have positive (negative) frequency in $t\;(t')$
on a spacelike surface in the future region $F$ ($P$) in
Fig.\ \ref{fig:hyperbolae}.
The Green's function derived below does satisfy this criterion and yields the
following explicit expression for the Wightman function in Gott space,
\begin{eqnarray}
\Delta^{(+)}(x,x') & = &
\left\langle 0+ \left| \phi(x) \phi(x') \right| 0- \right\rangle
\nonumber \\
& = & \int_0^{\infty}\,d\eta \sum_{n = - \infty}^{\infty}
\phi^{(+)}_{\eta, n}(x) \phi^{(-)}_{\eta, n}(x')\,,
\label{eq:Wightman:G}
\end{eqnarray}
where $\phi$ is the on mass shell eigenfunction defined in Eq.\
(\ref{eq:phi+}).

Because there is no time-ordering of the points in the acausal region $A$,
one cannot construct the time-ordered product, even given the Wightman
function, unless at least one of the points is outside this region.
That is, one cannot construct an expression like Eq.\ (\ref{eq:timeordered:M})
for Gott space.
One can, however, look for solutions to the Green's function equation
(\ref{eq:green's equation}); the solution to the equation is
\widetext
\begin{eqnarray}
G(x,x')
& = &
\int_0^{\infty}\,d\lambda \int_0^{\infty}\,d\eta \sum_{n = -\infty}^{\infty}
{
\psi^{(+)}_{\lambda,\eta,n}(x')^{\ast}
\psi^{(+)}_{\lambda,\eta,n}(x) +
\psi^{(-)}_{\lambda,\eta,n}(x')^{\ast}
\psi^{(-)}_{\lambda,\eta,n}(x)
\over - \lambda + m^2 -i \epsilon + (n\pi/2 + \alpha\eta)^2/Y_0^2}
\nonumber \\
& &  \null + \int^0_{- \infty}\,d\lambda \int_{-\infty}^{\infty}\,d\eta
\sum_{n = -\infty}^{\infty}
{\psi^{s}_{\lambda,\eta,n}(x')^{\ast}
\psi^{s}_{\lambda,\eta,n}(x)
\over - \lambda + m^2 + (n\pi/2 + \alpha\eta)^2/Y_0^2}  \,.
\label{eq:green's function:G}
\end{eqnarray}
\narrowtext\noindent
This expression is well defined by virtue of the $- i \epsilon$ in the
denominator which tells how to go around the pole as in the Minkowski case.
Its inclusion assures that the answer
satisfies the conditions given above as long as $x$ or $x'$ is in the causal
region.
The expression  (\ref{eq:green's function:G}) solves the Green's function
equation (\ref{eq:green's equation}) and satisfies the positive frequency
condition in the regions where it can be applied.
Since Green's functions can only differ
by a solution to the homogeneous equation, and, by the argument of
Sec.\ \ref{eigen}, a solution to the homogeneous equation is uniquely
determined by its value in the past and future regions, this must be the
unique Green's function which satisfies the positive frequency conditions in
the causal regions.

Alternatively, the functional integral definition \ref{eq:functional integral}
of the matrix element of the product of two field operators may
be used; however, it can no longer be derived from an operator formulation of
the theory.
No foliation of the spacetime using
spacelike Cauchy surfaces exists; hence there is no
global time parameter which can used to describe the time evolution of the
system, and, by the same token, there is no Hamiltonian which can be used to
generate the time evolution.
Thus, the matrix element cannot be written as a sum over a complete set of
states (field configurations) at each time.
However, one can imagine defining the quantum field theory by the functional
integral.
The non-trivial causal properties mean that the classical field theory also
has no Hamiltonian and no global time parameter.
This does not prevent one from defining the field theory and deriving its
equations of motion by varying an action defined over the space.
That procedure yields all the usual structures in the case of a causal space
and provides a generalization in the case of an acausal space.
The corresponding generalization of the quantum theory is to define the matrix
elements as the functional integral over field configurations with the complex
measure given in Eq.\ (\ref{eq:functional integral}).

If this procedure is adopted, the resultant propagator from
Eq.\ (\ref{eq:functional integral}),
\begin{eqnarray}
G(x, x') & = & i \int \left[d\phi\right]
e^{iW[\phi]} \phi(x) \phi(x') \,,
\end{eqnarray}
where, using the expansion, Eq.\ (\ref{eq:expand}),
\widetext
\ifpreprintsty
\begin{eqnarray}
W[\phi] & = & \sum_{\lambda, \eta, n}
\left[ m^2 - i \epsilon - \lambda + (n\pi/2 + \alpha\eta)^2/Y_0^2 \right]
\hspace{5em}
 \nonumber \\
 & &
\left[
\phi^{(+)}_n(\lambda, \eta)^{\ast} \phi^{(+)}_n(\lambda, \eta) +
\phi^{(-)}_n(\lambda, \eta)^{\ast} \phi^{(-)}_n(\lambda, \eta) +
\phi^{s}_n(\lambda, \eta)^{\ast} \phi^{s}_n(\lambda, \eta)
\right]/2\,,
\end{eqnarray}
\else
\begin{eqnarray}
\lefteqn{W[\phi] = \sum_{\lambda, \eta, n}
( m^2 - i \epsilon - \lambda + (n\pi/2 + \alpha\eta)^2/Y_0^2 )}\hspace{5em}
 \nonumber \\
 & &
\left(
\phi^{(+)}_n(\lambda, \eta)^{\ast} \phi^{(+)}_n(\lambda, \eta) +
\phi^{(-)}_n(\lambda, \eta)^{\ast} \phi^{(-)}_n(\lambda, \eta) +
\phi^{s}_n(\lambda, \eta)^{\ast} \phi^{s}_n(\lambda, \eta)
\right)/2\,;
\end{eqnarray}
\fi
\narrowtext\noindent
with the $\eta$ sum running over positive values for the $\phi^{(\pm)}$
terms and over all values for the $\phi^s$ terms.
The $\lambda$ sum runs over positive (negative) values for
$\phi^{(\pm)}$ ($\phi^{s}$).
The result of the integration is precisely that of
Eq.\ (\ref{eq:green's function:G}).
It is shown in Appendix \ref{Images} that
Eq.\ (\ref{eq:green's function:G})
is precisely equal to
the sum of images given in Eq.\ (\ref{eq:green-sum})
with the additional information that the
intervals $s(x, x'_n)$ are to be defined by the analytic continuation
\ifpreprintsty
\begin{equation}
s(x,x') = \sqrt{ (y-y')^2 + (x-x')^2 - (|t-t'| - i \epsilon)^2}\,,
\end{equation}
\else
\begin{eqnarray}
\lefteqn{s(x,x') =}
\hspace {1em} \nonumber \\
& & \sqrt{ (y-y')^2 + (x-x')^2 - (|t-t'| - i \epsilon)^2}\!,
\end{eqnarray}
\fi
with the requirement that the real part of the square root be positive.
As a result of this, the Green's function cannot be regarded as a function of
$x$ with $t$ given a fixed imaginary part; the different terms in the sum have
different imaginary parts for $t$ because $t$ may be greater than  $t'_n$ for
some terms and less than $t'_n$ for other terms.
However, $s(x,x')$ is an analytic function of $t$:
Suppose that $x'$ is in the acausal region and that $x$ is in the future
region.
Then all the terms in the sum may be defined by giving $t$ a negative
imaginary part.
The function is analytic in $t$ with branch points at
$ t'_n \pm \sqrt{(y - y'_n)^2 + (x - x'_n)^2}$.
As $t$ is continued past each of the branch points there is a
perfectly well defined continuation: $t$ goes below every branch point
with a positive square root and above every branch point with a
negative square root.
The result is that those terms which are associated with images to
the past (future) of $x$ have positive (negative) frequency in $t$.
Those terms which are associated with images which are spacelike with respect
to $x$ are independent of the choice of frequency.

The particles created by the operators propagate forward in time with
positive frequencies.
That is, if a field operator acts at $x'$ creating a particle, it can be
annihilated at any point within the future light cone of $x'$, and it carries a
positive frequency if that happens.
Of course, as in Minkowski space, operators at spacelike separations can
create and annihilate a particle.
This is because a positive frequency excitation cannot be localized, but
is spread out over the Compton wave length
of the particle with an amplitude which
decays exponentially with distance.
Those terms in the Green's function which correspond to a spacelike separation
between $x$ and the image of $x'$ decay exponentially with increasing
separation.

Although there is a perfectly good interpretation of the Green's function in
terms of the particles which are produced, there is no such interpretation in
terms of the ordering of the field operators.
As long as at least one point is outside the acausal region, there is a well
defined time ordering of the points, and the field operators may be
interpreted as having the corresponding order.
However,
when both points are in the acausal region, the fields cannot be ordered in
accord with the points; the ordering is that annihilation occurs to the future
of creation.
Thus the `ordering' is an ordering of propagation, not an ordering
of the field operators.
It is precisely this ordering which occurs when an interacting field theory is
considered.
This will be discussed further in the following section.

\section{Operator Products}
\label{stress}

The Green's function derived in Sec.\ \ref{Green} gives an explicit
representation for the vacuum matrix element of free-field operator products.
It may be written as
\begin{equation}
\left\langle \left( \phi(x) \phi(x') \right)_+ \right\rangle =
 -i \sum_{n = - \infty}^{\infty} G_0(x, x'_n) \,,
\end{equation}
where $x'_n$ is the position of the $n\mbox{th}$ image of the point $x'$
in the covering space.
If either $x$ or $x'$ are outside the acausal region, the ordering is a
time ordering.
Otherwise it is not an ordering of the field operators, but
rather an ordering of the propagation of the field excitations.

As $x' \rightarrow x$, the $n=0$ term in the sum diverges; this is just the
standard divergence of field theory in Minkowski space.
It is associated with various renormalizations which are
not of concern here and will be dropped.

The remaining terms are in general finite as the points approach each
other; in particular
\widetext
\begin{equation}
s(x, x'_n) \rightarrow s_n(x) = \left\{
\begin{array}{ll}
\sqrt{ (-x^+x^-)\sinh^2 n\alpha + (2nY_0)^2} \,,
& \qquad n \mbox{ even,}
\\
\sqrt{ (x^+e^{n\alpha} + x^-e^{- n\alpha})^2 + (2y + 2nY_0)^2} \,,
& \qquad n \mbox{ odd,}
\end{array}
\right.
\end{equation}
As a cursory examination of
Fig.\ \ref{fig:hyperbolae} will show,
the terms with odd $n$ are always spacelike,
and no choice of $n$ can make $s_n$ go to zero
except along the world lines of the
point masses, where $x^+ = - e^{\mp 2 \alpha}x^-$, $ y = \mp Y_0$.
The terms with even $n$ are spacelike for $x$ outside the acausal region.

The expectation value of $\phi^2$ can now be written as
\begin{equation}
\langle \phi^2(x) \rangle = - i \sum_{n = 1}^{\infty}
\left[ G_0(x,x_{n}) + G_0(x,x_{-n}) \right] \equiv -i \tilde G(x,x)\,.
\label{eq:renormalized Green}
\end{equation}
Each term is finite for $x$ in the causal regions.
The first question is the
convergence of the sum.
For non-zero mass, the sum is always convergent because
$-i G_0 \sim e^{-ms}/s^q$,
where $q$ is 1 for $2+1$ dimensions and $3/2$ for $3+1$ dimensions.
In either case, the exponential assures convergence of the sum.
In the zero mass case $ - i G_0 \sim 1/s^q$, where $q$ is 1 for $2+1$
dimensions, and 2 for $3+1$ dimensions.
Again, the sum converges because the asymptotic behavior of $s_n^{-q}$ is
$e^{-q n \alpha}$.

As the Cauchy horizon is approached, the convergence is more delicate.
The odd $n$ terms still have the same asymptotic behavior in $n$ as before
(except at the intersection of the Cauchy horizons $x^+ = 0 = x^-$).
The even $n$ terms still converge for non-zero mass
\begin{equation}
\left\langle \left( \phi^2(x) \right)_+ \right\rangle =
 \left\{
 \begin{array}{ll}
\displaystyle \sum_{n = 1}^{\infty}
 {\textstyle 2 \exp({- 4nY_0m}) \over \textstyle 4\pi 4nY_0}
 = {1\over 8\pi Y_0}\ln\left(1\over 1 - e^{-4 Y_0m}\right) \,,
 & \qquad\mbox{$2+1$ dim,}
 \\
 \noalign{\vspace{0.3ex}}
\displaystyle \sum_{n = 1}^{\infty}
 {\textstyle 2 m K_1(4nY_0m) \over \textstyle 2\pi^2 4nY_0} \,.
 & \qquad\mbox{$3+1$ dim,}
 \end{array}
\right.
\end{equation}
\narrowtext\noindent
In the massless case, the sum diverges as $x^+x^- \rightarrow 0$ in
$2+1$ dimensions but is finite in $3+1$ dimensions.
This divergence is logarithmic and therefore integrable; it appears for low
dimension and reflects the limited phase space available for the wave to
spread in.

In the acausal region the situation is more complicated.
For large $n$,
$s_n$ is imaginary with an infinitesimal positive real part which  assures
convergence just as in the causal region.
There are series of surfaces defined by
$x^+x^- = (4nY_0/\sinh 4n\alpha)^2$ on which the interval
$s_{2n}(x)$ vanishes.
These are precisely the polarized hypersurfaces \cite{Thorne:vacuum}
discussed in Section \ref{acausal}, where the point $x$ can be connected
to itself by a lightlike geodesic with winding number $n$.
The expectation value of $\phi^2(x)$ is singular on these surfaces since the
point $x$ is on its own lightcone.
Furthermore, the surfaces are dense as the
Cauchy horizon is approached;
that is, there are an infinite number of such
surfaces between every point in the acausal region and the Cauchy horizon.
The singularity is the standard lightcone singularity $s^{-q}$, where $q$ is
1 (2) in $2+1$ ($3+1$) dimensions.
An imaginary $s_n(x)$ indicates that $x$ is within its own future
lightcone (for paths with winding number $n$), and, therefore, a particle can
be created at $x$ to propagate forward until it is annihilated at
the same point.
The propagator is then complex with a positive phase (because of the positive
frequency condition).
This will be important in determining the properties of an interacting field in
Section \ref{interacting}.

\widetext
The stress energy of the field is given by
\begin{equation}
T_{\mu\nu} = \phi(x)_{,\mu}\phi(x)_{,\nu} - \eta_{\mu\nu}
(1/2)\left[\phi(x)^{,\lambda}\phi(x)_{,\lambda}  + m^2 \phi^2(x)\right]\!.
\end{equation}
To compute its matrix elements, note that derivatives of the function
\begin{equation}
s_{2n}(x,x') =
\sqrt{(x^+ - e^{4n\alpha} x'^+)(x^- - e^{- 4n\alpha} x'^-)
+ ( y - y' - 4n Y_0)^2}
\end{equation}
are given by
\begin{eqnarray}
{\partial s_{2n}(x,x') \over \partial x^{\pm} } =
{(x^{\mp} - e^{\mp 4n \alpha}x'^{\mp})\over 2 s_{2n}(x,x')} \,,
 & \qquad &
{\partial s_{2n}(x,x') \over \partial x'^{\pm} } =
{(x'^{\mp} - e^{\pm 4n \alpha}x^{\mp})\over  2 s_{2n}(x,x')} \,,
\nonumber \\
{\partial s_{2n}(x,x') \over \partial y } =
{(y - y' - 4nY_0)\over s_{2n}(x,x')} \,,
 & \qquad &
{\partial s_{2n}(x,x') \over \partial y' } =
 {- (y - y' - 4nY_0)\over s_{2n}(x,x')} \,,
\end{eqnarray}
which imply that
\begin{eqnarray}
\left.
\partial^2 f(s_{2n}(x,x')) \over \partial x^{\pm}\partial x'^{\pm}
\right|_{x'=x}
& = &
 - \left[ (x^{\mp})^2\sinh^2 2n\alpha\over s_{2n} \right]
\left[ f'(s_{2n})\over s_{2n}\right]' \!,
\nonumber \\
\left.
\partial^2 f(s_{2n}(x,x')) \over \partial x^{\pm}\partial y'
\right|_{x'=x}
& = &
\left[ \pm (4nY_0) e^{\mp 2n\alpha}x^{\mp}\sinh 2n\alpha\over s_{2n} \right]
\left[ f'(s_{2n})\over s_{2n}\right]' \!,
\nonumber \\
\left.
\partial^2 f(s_{2n}(x,x')) \over \partial y \partial x'^{\pm}
\right|_{x'=x}
& = &
\left[ \pm (4nY_0) e^{\pm 2n\alpha}x^{\mp}\sinh 2n\alpha\over s_{2n} \right]
\left[ f'(s_{2n})\over s_{2n}\right]' \!,
\nonumber \\
\left.
\partial^2 f(s_{2n}(x,x')) \over \partial x^{\pm}\partial x'^{\mp}
\right|_{x'=x}
& = &
\left[ x^{\mp}x^{\pm}e^{\mp4n\alpha}\sinh^2 2n\alpha\over s_{2n} \right]
\left[ f'(s_{2n})\over s_{2n}\right]'
\\
& &
\null - \left[e^{\mp 4n \alpha}\over 2 \right]
\left[f'(s_{2n}(x,x'))\over s_{2n}(x,x')\right] \!,
\nonumber \\
\left.
\partial^2 f(s_{2n}(x,x')) \over \partial y \partial y'
\right|_{x'=x}
& = &
 - \left[ (4nY_0)^2 \over s_{2n} \right]
\left[ f'(s_{2n})\over s_{2n}\right]'
\nonumber \\
& &
\null - \left[f'(s_{2n}(x,x'))\over s_{2n}(x,x')\right] \!,
\nonumber
\end{eqnarray}
which in turn yield
\begin{eqnarray}
\lefteqn{
\left.
\partial^{\lambda}\partial'_{\lambda} f(s_{2n}(x,x')) + m^2 f(s_{2n})
\right|_{x' = x} =\ (\cosh 4 n \alpha - 1)} \hspace{ 5em}
\nonumber \\
& &
\left[ {4 x^+x^- \sinh^2 2n\alpha \over s_{2n}^2}
\left( m^2 f(s_{2n}) - {3f'(s_{2n})\over s_{2n}} \right)
- {2\over s_{2n}} f'(s_{2n}) \right]\!.
\end{eqnarray}
Moreover, for $f(s_{2n}(x,x')) = G(x,x'_{2n})$, it  obeys the wave equation
\begin{equation}
0 = (-\partial^2 + m^2)f(s) =
-3 \left( f'(s)\over s\right)
- s \left( f'(s)\over s\right)' + m^2 f(s)\,,
\end{equation}
where $f(s_{2n}) = -i G_0(x,x_n)$.

In $2 + 1$ dimensions the matrix elements of the stress tensor are therefore
given by
\begin{eqnarray}
\left\langle T^{\pm\pm}(x) \right\rangle
& = &
\sum_{n = 1}^{\infty}
\left[ - 8 x^{\pm}x^{\pm}\sinh^2 2n\alpha\over s^2_{2n} \right]
\left[ m^2 f(s_{2n}) - {3f'(s_{2n})\over s_{2n}} \right] \!,
\nonumber \\
\left\langle T^{\pm y}(x) \right\rangle
& = &
0 \,,
%\sum_{n = 1}^{\infty} \left[
%\mp 4 x^\pm (4nY_0) \sinh 2n\alpha\cosh 2n\alpha\over s^2_{2n} \right)
%\left( m^2 f(s_{2n}) - {3f'(s_{2n})\over s_{2n}} \right) \,,
\nonumber \\
\left\langle T^{\pm \mp}(x) \right\rangle
& = &
\sum_{n = 1}^{\infty}
2\left[ 4 x^{+}x^{-}\sinh^2 2n\alpha \over s^2_{2n} \right]
\left[ m^2 f(s_{2n}) - {3f'(s_{2n})\over s_{2n}} \right]
- 4 {f'(s_{2n}) \over s_{2n} } \!,
\nonumber \\
\left\langle T^{yy}(x) \right\rangle
& = &
 \sum_{n = 1}^{\infty}
\left[ - 2 (4nY_0)^2 \over s^2_{2n} \right]
\left[ m^2 f(s_{2n}) - {3 f'(s_{2n}) \over s_{2n}} \right]
- 2 {f'(s_{2n}) \over s_{2n} }
\\
& & \null - (\cosh 4 n \alpha - 1)
\left\{ {4 x^+x^- \sinh^2 2n\alpha \over s_{2n}^2}
\left[ m^2 f(s_{2n}) - {3 f'(s_{2n}) \over s_{2n}} \right]
- 2 {f'(s_{2n})\over s_{2n}} \right\}\!,
\nonumber
\end{eqnarray}

For $m \neq 0$ and $x^+x^- \neq 0$, these sums are convergent, yielding
a finite stress energy tensor at all points in the causal regions.
As $m \rightarrow 0$, the convergence provided by the exponential $e^{-ms}$
disappears, but the sums still converge because $f'/s \sim 1/s^3$, and the
asymptotic behavior in $n$ is $e^{-2n\alpha}$ leading to convergence.
As $x^+x^- \rightarrow 0$ the sums are more delicate.
For $m \neq 0$ and $x^+x^- = 0$, $s_{2n} = 4nY_0$, and the asymptotic behavior
of the terms in the sums is $\sim e^{4n(\alpha - mY_0)}$; the sum converges
provided $\alpha < mY_0$.
Thus, for sufficiently small relative rapidity of
the mass points, the stress tensor is regular on the Cauchy horizons.
For $m = 0$ the sums diverge on the Cauchy horizons.
The sums, which are somewhat tricky to evaluate for large rapidity $\alpha$,
are easily estimated for small $\alpha$.  In that case they are of the form
\begin{equation}
S = \sum_n {e^{4n\alpha}n^q \over [(-x^+x^-) e^{4n\alpha} + n^2(4Y_0)^2]^p} \,,
\end{equation}
which may be approximated as an integral over $n$,
and then evaluated by using
a steepest descent approximation.
The result is
\begin{equation}
S \sim {\bar n^q \over (-x^+x^-) ((4Y_0)^2 \bar n^2)^{(p-1)}} \,,
\end{equation}
where $\bar n \sim \ln ( (4Y_0)^2 / (-x^+x^-) )$.
The matrix elements of the stress-energy tensor are then, as the Cauchy
horizons are approached,
\begin{eqnarray}
\left\langle T^{\pm\pm}(x) \right\rangle
& \sim &
{ - 2 x^{\pm}x^{\pm} \over
(-x^+x^-) \left[(4Y_0)^2 \ln^2 ( (4Y_0)^2/(-x^+x^-)) \right]^{3/2} }  \,,
\nonumber \\
\left\langle T^{\pm y}(x) \right\rangle
& \sim &
0 \,,
\nonumber \\
\left\langle T^{\pm \mp}(x) \right\rangle
& \sim &
\mbox{finite} \,,
\label{eq:stress}
\\
\left\langle T^{yy}(x) \right\rangle
& \sim &
{ 1 \over 2 (-x^+x^- )
\left[ (4Y_0)^2 \ln^2 ( (4Y_0)^2/(-x^+x^-)) \right]^{1/2} } \,.
\nonumber
\end{eqnarray}
\section{Interacting Fields}
\label{interacting}

For an interacting scalar field, the action
(\ref{eq:action})
of the free field is replaced by
\begin{equation}
W[\phi] = - \frac{1}{2}\int\,d^4x
\left[
(\partial\phi)^2
+(m^2 - i \epsilon)\phi^2 + {\lambda\over 4!} \phi^4
\right]\!.
\end{equation}
When this is used in the functional integral
expression for the propagator
\begin{equation}
\Delta_F(x,x') =  i \int\,\left[d\phi\right] e^{iW[\phi]} \phi(x)\phi(x') \,,
\end{equation}
the resultant expression for $\Delta_F$ to first order in the coupling constant
$\lambda$ is,
after renormalization,
\begin{equation}
\Delta_F(x,x') = G(x,x') +
i {\lambda \over 2} \int\,dy G(x,y) \tilde G(y,y) G(y,x')\,,
\end{equation}
where $\tilde G$, as defined in Eq.\ (\ref{eq:renormalized Green}),
denotes the Green's function with the Minkowski space, zero
winding number term, removed.
This is non-singular as the points approach each other, and the dropped term
simply renormalizes the mass of the scalar particle.

If $x'$ is in one of the causal regions and in the past of $x$, then $\Delta_F$
may be integrated over a spacelike hyperboloid with $\phi^{(+)}_{\eta,n}(x')$,
and the reduction formula, (\ref{eq:reduction}), together with the
orthonormality relation, (\ref{eq:phi orthonormality}), yields the
result
\begin{equation}
\langle 0 | \phi(x) | \eta,n ; \sigma(x')\rangle =
\phi^{(+)}_{\eta, n}(x) + i {\lambda \over 2} \int_{y > \sigma(x')}\,dy
G(x,y) \tilde G(y,y) \phi^{(+)}_{\eta, n}(y) \,,
\end{equation}
where $\sigma(x')$ denotes the spacelike surface over which the integral was
done, and $y > \sigma(x')$ means that the integral is restricted to points in
the future of $\sigma(x')$.

Applying the reduction formula (\ref{eq:reduction}) again with
${\phi^{(+)}}^*$, the matrix element becomes
\begin{eqnarray}
\lefteqn{\langle \eta_1, n_1 ; \sigma(x) | \eta,n ; \sigma(x')\rangle
 = } \hspace { 5em} \nonumber \\
 & &
\delta_{n_1,n}\delta(\eta_1 - \eta) -
i {\lambda \over 2} \int_{\sigma(x) > y > \sigma(x')}\,dy
\phi^{(+)}_{\eta_1, n_1}(y)^* V(y) \phi^{(+)}_{\eta, n}(y) \,,
\label{eq:scatt}
\end{eqnarray}
\narrowtext\noindent
where $V(y) \equiv - i\tilde G(y,y)$.
As was discussed in Sec.\ \ref{stress}, $V(y)$ is real for $y$ in the causal
regions.
As long as the region of integration does not include any of the acausal
region, the additional contribution to the matrix element is purely imaginary
and only contributes terms of order $\lambda^2$ to the unitarity relation.
But  $V(y)$ is complex in the acausal region,
and the matrix element contains a real part coming from the integration
over that region.
As a result, the eigenvalues of the scattering matrix (\ref{eq:scatt}) are
not associated with real phases.
Furthermore, the imaginary parts of the phases
have no definite sign so that some probabilities will be greater than 1 (but
of first order in $\lambda$), and
the failure of unitarity cannot be associated with a failure to include all
possible final states.
Unitarity, expressed in terms of the particles which start in the past
causal region and end in the future causal region, fails because
real (rather than virtual) particles are
created in the acausal region, are propagated around a closed time-like path,
and are annihilated at the spacetime point at which they were created.
If one could treat the terms with different winding numbers as physically
distinct events so that there was no interference between them, there would be
no problem with unitarity:
The specification of the state would include a specification of whether an
on-shell particle was created and, if so, what its winding number was.
Then, the sum over states would include the sum over winding numbers and the
phase would be a kind of final state phase which would cancel in the sum over
final states.
The interference term between no production and production with some winding
number would not enter and, in this approximation, there would be no lack of
unitarity.
Since the on-shell particles
appear in neither the future causal region nor the past causal
region, there
is no way to specify, in terms of data on the initial or final surfaces,
what on-shell
particles were created with what winding numbers never to emerge from the
acausal region.

Deutsch \cite{Deutsch} has discussed this problem by describing the behavior of
the system in the acausal region by means of a density matrix.
Although the density matrix does provide partial information about what
happens in the acausal region and is precisely the kind of information that is
required in order to address the unitarity problem, it is additional
information imposed from the outside (subject to some consistency conditions)
and does not arise naturally from the theory.

The unitarity problem is directly associated with the absence of a global
Cauchy surface.
If there were such a surface one could specify data or the
state of the system on that surface, and the propagation would uniquely
determine the state for the entire spacetime.
Here, the data on spacelike surfaces in the causal regions completely
determines the non-interacting field but not the interacting quantum field.
(The spacelike surface is not a global Cauchy surface.)

To summarize,
the potential $V$ in Eq.\ (\ref{eq:scatt}) may be written as a sum of terms
arising from paths with different winding numbers.
The contribution of each term to the amplitude
contains information about a real particle which
was created, then annihilated at the same spacetime point
after winding around the singularities $n$ times;
but the specification of the state does not include the information about what
particles were produced.
It is hard to see how the specification of the state could be enlarged to
include such information: The real potential which arises from the virtual
creation and annihilation of particles cannot and should not be separated into
contributions with different winding numbers, and one expects that the analytic
continuation from virtual to real production to be valid here as well as in
causal spaces.
If one could extend the surface on which the data is specified into the
acausal region, one could then expect to have a unitarity relation in which the
probabilities added to unity.
However, there is no surface in the acausal region on
which unrestricted data may be specified, and no way to separate the
contributions from different winding numbers so that they do not interfere
with the zero winding number contribution.
The inclusion of higher order corrections in $\lambda$ cannot save the
situation because the failure is first order in $\lambda$.
Furthermore,
the work of Friedman {\it et al.\/} \cite{Friedman:unitarity} shows similar
results in order $\lambda^2$; the theory is not unitary by any usual standard.

\ifpreprintsty\pagebreak\fi
\section{Back Reaction}
\label{back}

The matrix element of the stress tensor is of the form Eq.\ (\ref{eq:stress})
for the divergent parts.
These terms are invariant under boosts in the $(x,t)$ plane and reflections in
$y$, and they are independent of $y$.
The most general metric in the $2+1$ dimensional space which satisfies these
symmetries is
\begin{equation}
ds^2 = e^{2\psi(\zeta)} dx^+ dx^- + e^{2\phi(\zeta)}dy^2\,,
\end{equation}
where $\zeta = - x^+x^-$.

The Einstein tensor for this metric is
\begin{equation}
\begin{array}{rcl}
G^{\pm\pm} & = &
-4 x^{\pm}x^{\pm} \left(\phi'e^{(\phi - 2 \psi)}\right)'e^{-\phi}\,,
\\
G^{\pm y} & = &
0\,,
\\
G^{\pm\mp} & = &
4 \left(- \zeta\phi'e^{\phi }\right)'e^{-(\phi + 2 \psi)}\,,
\\
G^{yy} & = &
4 \left(- \zeta\psi'\right)'e^{-2\psi}\,.
\end{array}
\end{equation}

The expression for the stress energy must be equal to this to lowest order;
hence,
\begin{equation}
\begin{array}{rcl}
\left(- \zeta\psi'\right)' & \sim &
{G \over 2 \zeta
\left[ (4Y_0)^2 \ln^2 [ (4Y_0)^2/\zeta] \right]^{1/2} }\!,
\\
\noalign{\vspace{0.2ex}}
\phi'' & \sim &
{2G  \over
\zeta \left[(4Y_0)^2 \ln^2 [ (4Y_0)^2/\zeta] \right]^{3/2} }\!.
\label{eq:sing}
\end{array}
\end{equation}
The $\pm\mp$ equation is non-singular at $\zeta \sim 0$, and it does not
provide any additional information.
The coefficients of the right sides of Eqs.\ (\ref{eq:sing}) are nontrivial
functions of $\alpha$ which cannot be determined by the methods used here;
however,
Newton's constant $G$ is included so that the units come out correctly.
In $3+1$ dimensions, the powers of $1/2$ and $3/2$ become $1$ and $2$
respectively because the propagator is, for $m =0$, $1/s^2$ rather than $1/s$.
The solutions to the equations are then
\widetext
\begin{equation}
\begin{array}{rcl}
\psi(\zeta)
& \sim &
\left\{
\begin{array}{ll}
- C(G/Y_0)\ln((4Y_0)^2/\zeta) (1 + \ln\ln((4Y_0)^2/\zeta)) \,,&
\mbox{\quad$2+1$ dim,}
\\
- C(G/Y_0^2) \ln\ln((4Y_0)^2/\zeta) \,,&
\mbox{\quad$3+1$ dim,}
\end{array}
\right.
\\
\phi(\zeta)
& \sim &
\mbox{finite,}
\end{array}
\end{equation}
where $C$ is a positive constant depending upon the dimension and
upon $\alpha$.

Since $\psi \sim - \infty$ as $\zeta \sim 0$,
the resultant metric is singular at the Cauchy horizons (the
coefficient of $dx^+dx^-$, $e^{2\psi}$, vanishes there).
The calculation of the stress tensor as modified by the change in the metric
is beyond the scope of this paper.

\acknowledgments

I would like to thank J. Friedman, J. Hartle, and K. Thorne for helpful
discussions on the properties of acausal spaces and on quantum field theories
in such spaces.
I would also like to thank L. Brown for many helpful comments about the
manuscript.
This work was supported in part by DOE grant DE-FG06-91ER40614.

\ifpreprintsty\pagebreak\fi
\appendix{Orthonormality}
\label{app:orthonormal}

The functions defined in Eq.\ (\ref{eq:psi+}),
\begin{eqnarray}
\psi_{w',\eta,n}^{(+)}(x^+, x^-, y) & = &
 {1\over (2\pi)^{3/2}}\frac{1}{2} \int^{\infty}_0\,dk^+
\int^{\infty}_{-\infty}\,dk^-
\delta( k^+ k^- + w') e^{i(k^+x^- + k^-x^+)/2} \nonumber \\
 & &  \qquad \left[
{e^{i{y\over Y_0}({n\pi\over2} + \alpha\eta)} \over \sqrt{2 Y_0}i^n}
\left(k^+\over - k^-\right)^{i\eta/2} +
 {e^{- i{y\over Y_0}({n\pi\over2} + \alpha\eta)} \over \sqrt{2 Y_0}i^{-n}}
\left(k^+\over - k^-\right)^{-i\eta/2} \right]\!,
\end{eqnarray}
Eq.\ (\ref{eq:psi-}),
\begin{eqnarray}
\lefteqn{\psi_{w',\eta,n}^{(-)}(x^+, x^-, y)  =
\left(\psi_{w',\eta,n}^{(+)}(x^+, x^-, y)\right)^{\ast} = } \qquad
\nonumber \\
 & & \qquad{1\over (2\pi)^{3/2}} \frac{1}{2}\int^{\infty}_0\,dk^+
\int^{\infty}_{-\infty}\,dk^-
\delta( k^+ k^- + w') e^{-i(k^+x^- + k^-x^+)/2} \nonumber \\
 & & \qquad \qquad\left[
{e^{i{y\over Y_0}({n\pi\over2} + \alpha\eta)} \over \sqrt{2 Y_0}i^n}
\left(k^+\over - k^-\right)^{i\eta/2} +
{e^{- i{y\over Y_0}({n\pi\over2} + \alpha\eta)} \over \sqrt{2 Y_0}i^{-n}}
\left(k^+\over - k^-\right)^{-i\eta/2}
\right]\!,
\end{eqnarray}
and Eq.\ (\ref{eq:psi-s}),
\begin{eqnarray}
\psi_{ - v',\eta,n}^{s}(x^+, x^-, y) & = &
 {1\over (2\pi)^{3/2}} \frac{1}{2}\int^{\infty}_0\,dk^+
\int^{\infty}_{-\infty}\,dk^-
\delta( k^+ k^- - v')  \nonumber \\
 & &  \hspace{- 5em} \left[
{e^{i(k^+x^- + k^-x^+)/2}
e^{i{y\over Y_0}({n\pi\over2} + \alpha\eta)} \over \sqrt{2 Y_0}i^n}
\left(k^+\over k^-\right)^{i\eta/2} \right. \qquad \\
 & + & \left. {e^{-i(k^+x^- + k^-x^+)/2}
 e^{- i{y\over Y_0}({n\pi\over2} + \alpha\eta)} \over \sqrt{2 Y_0}i^{-n}}
\left(k^+\over k^-\right)^{-i\eta/2} \right]\!, \nonumber
\end{eqnarray}
form a complete orthonormal set, where $w$ and $v$ are both positive.
To see this, calculate the inner product over the physical space
\begin{equation}
I(w',\eta',n',a' ;w,\eta,n,a ) =
\frac{1}{2}\int^{\infty}_{-\infty}\,dx^+
\int^{\infty}_{-\infty}\,dx^-
\int^{Y_0}_{-Y_0}\,dy
\psi^{a'}_{w',\eta',n'}(x)
\psi^{a}_{w,\eta,n}(x)\,.
\label{eq:inner}
\end{equation}
For each combination of functions, the integrations over $(x^+,x^-)$ yield a
delta function of the momenta $(k^+,k^-)$, which in turn forces the momenta in
the definitions of the $\psi$'s to be equal.
The integral thus vanishes
unless $a' = a$ (otherwise the ranges of the $k$ integrations do not overlap).
Since the two momenta are equal, a factor of $\delta(w' - w)$
appears.
Hence
\begin{equation}
I(w',\eta',n',a' ;w,\eta,n,a ) =
\delta_{a',a} \delta(w' - w)
I^a(w ; \eta',n' ; \eta,n ),
\end{equation}
where
\begin{equation}
I^a(w ; \eta',n' ; \eta,n ) =
\int^{Y_0}_{-Y_0}\,dy
\int_0^{\infty}\,dk^+ \int_{-\infty}^{\infty}\,dk^- \delta(k^+k^- + w)
E^a_{w}(y; \eta', n', \eta, n ) ,
\label{eq:E}
\end{equation}
and
\begin{eqnarray}
\lefteqn{E^a_{w}(y; \eta', n', \eta, n ) = } \qquad \nonumber \\
& & \left\{
    \begin{array}{lr}
    \begin{array}{c}
    \bigg[
    \left( e^{i(y/ Y_0)\left({(n - n')\pi/ 2} + \alpha(\eta - \eta')\right)}
    / 8\pi Y_0 i^{n - n'} \right)
    \left( k^+/ - k^- \right)^{i(\eta - \eta')/2} \\
    \null + (n',\eta')\rightarrow ( -n',-\eta')\bigg]
    + \bigg[(n,\eta)\rightarrow ( -n,-\eta)\bigg] \,,
    \end{array} & a = \pm\,, \\
    \begin{array}{c}
    \bigg[\left( e^{i(y/ Y_0)\left({(n - n')\pi/ 2} + \alpha(\eta -
\eta')\right)}
    / 8\pi Y_0 i^{n - n'} \right)
    \left( k^+/ k^- \right)^{i(\eta - \eta')/2}\bigg] \\
    \null +
    \bigg[(n',\eta', n, \eta )\rightarrow ( -n',-\eta', -n , -\eta )\bigg]
    \,,
    \end{array} & a = s\,.
    \end{array}
    \right.
\end{eqnarray}

The integrations over $k^{\pm}$ now yield $2\pi\delta(\eta \mp \eta')$, with
the terms with the plus sign appearing only for $a = \pm$, in which case they
vanish because $\eta > 0$.
The integral becomes
\begin{equation}
I(w',\eta',n',a' ;w,\eta,n,a ) =
\delta_{a',a} \delta(w' - w)\delta(\eta' - \eta)
\int^{Y_0}_{-Y_0}\,dy E^a_{w, \eta}(y, n', n ) ,
\end{equation}
where
\begin{equation}
E^a_{w, \eta}(y, n', n ) =
 \left( e^{i(y/ Y_0)\left({(n - n')\pi/ 2} \right)}
    / 4 Y_0 i^{n - n'}\right)
    + (n, n')\rightarrow ( -n,- n') \,.
\end{equation}
The integration over $y$ then yields $\delta_{n',n}$ and,
\begin{equation}
I(w',\eta',n',a' ;w,\eta,n,a ) =
\delta_{a',a} \delta(w' - w)\delta(\eta' - \eta)
\delta_{n',n}.
\end{equation}
The functions thus form an orthonormal set.

Completeness is established by calculating
\begin{eqnarray}
\lefteqn{I(x'^+, x'^-, y'; x^+, x^-, y) = }\qquad \nonumber \\
&\qquad &
\int_0^{\infty}\,dw \int_0^{\infty}\,d\eta \sum_{n = -\infty}^{\infty}
\left[
\psi^{(+)}_{w,\eta,n}(x')^{\ast}
\psi^{(+)}_{w,\eta,n}(x) +
\psi^{(-)}_{w,\eta,n}(x')^{\ast}
\psi^{(-)}_{w,\eta,n}(x) \right] \nonumber \\
&\qquad &
\null + \int^0_{- \infty}\,dw \int_{-\infty}^{\infty}\,d\eta
\sum_{n = -\infty}^{\infty}
\psi^{s}_{w,\eta,n}(x')^{\ast}
\psi^{s}_{w,\eta,n}(x) \,.
\label{eq:complete1}
\end{eqnarray}
The sums over $n$ are of the form
\begin{equation}
\sum^{\infty}_{n = - \infty} { e^{in\pi ( y - y')/2}\over 4Y_0} =
\delta( y - y')\,,
\end{equation}
or
\begin{equation}
\sum^{\infty}_{n = - \infty} { e^{in\pi ( y + y' + 2 Y_0)/2}\over 4Y_0} =
\delta( y + y' + 2 Y_0 )\,;
\end{equation}
the second form vanishes for $ -Y_0 < (y,y') < Y_0$.
Hence the sum over $n$ yields
\begin{eqnarray}
\lefteqn{\int_0^{\infty}\,d\eta \sum_{n = -\infty}^{\infty}
\left[
\psi^{(+)}_{w,\eta,n}(x')^{\ast}
\psi^{(+)}_{w,\eta,n}(x) +
\psi^{(-)}_{w,\eta,n}(x')^{\ast}
\psi^{(-)}_{w,\eta,n}(x) \right] } \hspace{ 15em} \nonumber \\
& = & \delta( y' - y) I^{\pm}(w; x'^+, x'^-; x^+, x^- )\,,
\label{eq:complete2}
\end{eqnarray}
where
\begin{eqnarray}
\lefteqn{ I^{\pm}(w, x'^+, x'^-, x^+, x^- ) } \qquad \nonumber \\
& = & \int_0^{\infty}\,d\eta {1\over (2\pi)^{3}}
\frac{1}{2}\int_0^{\infty}\,dk^+\int_{-\infty}^{\infty}\,dk^-
\frac{1}{2}\int_0^{\infty}\,dk'^+\int_{-\infty}^{\infty}\,dk'^-
\delta( k^+k^- + w ) \delta( k'^+k'^- + w)
\nonumber \\
& &\qquad \left[e^{i(k^+x^- + k^-x^+ - k'^+x'^- - k'^-x'^+)/2} +
e^{- i(k^+x^- + k^-x^+ - k'^+x'^- - k'^-x'^+)/2}\right]
\nonumber \\
& & \qquad \left[(k^+k'^-/k^-k'^+)^{i\eta/2} +
(k^+k'^-/k^-k'^+)^{- i\eta/2} \right]
\label{eq:complete3}
\\
& = & {1\over (2\pi)^{2}}
\frac{1}{2}\int_{0}^{\infty}\,dk^+\int_{-\infty}^{\infty}\,dk^-
\frac{1}{2}\int_{0}^{\infty}\,dk'^+\int_{-\infty}^{\infty}\,dk'^-
\nonumber \\
& & \qquad \delta( k^+k^- + w ) \delta( k'^+k'^- - k^+k^-)
\delta( \ln(k^+k'^-/k^-k'^+)^{1/2})
\nonumber \\
& & \qquad \left[ e^{i(k^+x^- + k^-x^+ - k'^+x'^- - k'^-x'^+)/2} +
e^{- i(k^+x^- + k^-x^+ - k'^+x'^- - k'^-x'^+)/2}  \right )  \nonumber \\
& = & {1\over (2\pi)^{2}}
\frac{1}{4}\int_{-\infty}^{\infty}\,dk^+\int_{-\infty}^{\infty}\,dk^-
\delta( k^+k^- + w )
\left[ e^{i(k^+(x^- -  x'^-) + k^-(x^+ - x'^+))/2} \right]\!.
\nonumber
\end{eqnarray}
A similar argument leads to the result
\begin{eqnarray}
\lefteqn{\int _{-\infty}^{\infty}\,d\eta \sum_{n = -\infty}^{\infty}
\psi^{s}_{w,\eta,n}(x')^{\ast}
\psi^{s}_{w,\eta,n}(x) =
 \delta( y' - y) I^{s}(w; \eta, x'^+, x'^-; x^+, x^- ) } \qquad
\label{eq:complete4}
\\
& = & \delta( y' - y) {1\over (2\pi)^{2}}
\frac{1}{4}\int_{-\infty}^{\infty}\,dk^+\int_{-\infty}^{\infty}\,dk^-
\delta( k^+k^- + w )
\left[ e^{i(k^+(x^- -  x'^-) + k^-(x^+ - x'^+))/2} \right]\!.
\nonumber
\end{eqnarray}
If Eqs.\ (\ref{eq:complete1}), (\ref{eq:complete2}), (\ref{eq:complete3}), and
(\ref{eq:complete4}) are combined the result
\begin{eqnarray}
\lefteqn{I(x'^+, x'^-, y'; x^+, x^-, y) }\qquad \nonumber \\
&\qquad = &
\int_0^{\infty}\,dw \int_0^{\infty}\,d\eta \sum_{n = -\infty}^{\infty}
\left[
\psi^{(+)}_{w,\eta,n}(x')^{\ast}
\psi^{(+)}_{w,\eta,n}(x) +
\psi^{(-)}_{w,\eta,n}(x')^{\ast}
\psi^{(-)}_{w,\eta,n}(x) \right] \nonumber \\
&\qquad &
\null + \int^0_{- \infty}\,dw \int_{-\infty}^{\infty}\,d\eta
\sum_{n = -\infty}^{\infty}
\psi^{s}_{w,\eta,n}(x')^{\ast}
\psi^{s}_{w,\eta,n}(x)
\nonumber \\
& \qquad = &  2 \delta(y' - y) \delta(x'^+ - x^+ ) \delta(x'^- - x^- )\,
\end{eqnarray}
is obtained, and completeness is established.

In order to calculate the functional integral in Sec.\ \ref{Green}, the
integral
\begin{eqnarray}
\lefteqn{I'(w',\eta',n',a' ;w,\eta,n,a ) = } \hspace{ 5em}
\nonumber \\
& & \frac{1}{2}\int^{\infty}_{-\infty}\,dx^+
\int^{\infty}_{-\infty}\,dx^-
\int^{Y_0}_{-Y_0}\,dy
\partial^{\mu}\psi^{a'}_{w',\eta',n'}(x)
\partial_{\mu}\psi^{a}_{w,\eta,n}(x)\,
\end{eqnarray}
is required.
The terms with derivatives with respect to $x^{\pm}$ yield
integrals which are the same as (\ref{eq:inner}) with extra factors of
$k^{\pm}$ inside the integrals which, after the $x^{\pm}$ integrations are
done, simply become a factor of $ -w$ multiplying (\ref{eq:inner}).
The term with derivatives with respect to $y$ yields
an integral which is again the same but with an extra factor of
$(n\pi/2 + \alpha\eta)(n'\pi/2 + \alpha\eta')/Y_0^2$;
this factor does not affect the previous arguments, and the result is an
overall factor of $(n\pi/2 + \alpha\eta)^2/Y_0^2$ in (\ref{eq:inner}).
Thus,
\begin{eqnarray}
\lefteqn{I'(w',\eta',n',a' ;w,\eta,n,a ) = } \hspace{ 5em}
\nonumber \\
& & \left[ - w + \left(n\pi/2 + \alpha\eta\right)^2/Y_0^2 \right]
\delta_{a',a}\delta_{n',n}\delta(\eta' - \eta)\delta(w' - w)\,.
\label{eq:lagrangian integral}
\end{eqnarray}

The eigenfunctions $\psi$ may be expressed in terms of Bessel functions
of imaginary order: The $k^{\pm}$ integrations may be done explicitly in
(\ref{eq:psi+}); for $w > 0 $, first calculate
\begin{eqnarray}
\lefteqn{
I(x, \eta) =
\int_0^{\infty}\,dk^+\int_{-\infty}^{\infty}\,dk^- \delta(k^+k^- +w)
\left(k^+ \over - k^-\right)^{i\eta/2} e^{i(k^+x^- + k^-x^+)/2}
} \hspace{ 5em}
\nonumber \\
& = &
\int_0^{\infty}\left(dk^+ \over k^+ \right)
\left(k^+ \over \sqrt{w}\right)^{i\eta}
e^{i(k^+x^- - w x^+/k^+)/2} \,.
\label{eq:Idef}
\end{eqnarray}
For $x^{\pm} > 0$, the change of variables
$k^+ \rightarrow \sqrt{w x^+/x^-} e^{\sigma}$
followed by a translation of $\sigma$ by $i\pi/2$ leads to the expression
\begin{eqnarray}
I(x, \eta)
& = &
\int_{-\infty}^{\infty}\,d\sigma
\left(x^+ \over x^-\right)^{i\eta/2} e^{i\eta\sigma}
e^{i\sqrt{w x^+ x^-}\sinh\sigma}
\nonumber \\
& = &
\int_{-\infty}^{\infty}\,d\sigma
\left(x^+ \over x^-\right)^{i\eta/2} e^{- \pi\eta /2} e^{i\eta\sigma}
e^{ - \sqrt{w x^+ x^-}\cosh\sigma}
\nonumber \\
& = &
2 \left(x^+ \over x^-\right)^{i\eta/2} e^{- \pi\eta /2}
K_{i\eta}\left(\sqrt{w x^+ x^- }\right)\!.
\end{eqnarray}
The original expression (\ref{eq:Idef})
is analytic for $ \mp \mbox{Im} x^{\pm} > 0$; hence
the value for all $x$ may be obtained by analytic continuation with the
result
\begin{equation}
I(x, \eta) = \left\{%
\begin{array}{ll}
2 \left(x^+ \over x^-\right)^{i\eta/2} e^{- \pi\eta /2}
K_{i\eta}\left(\sqrt{w x^+ x^- }\right)\!,
& \qquad x^{\pm} > 0\,,
\\
2 \left(- x^+ \over - x^-\right)^{i\eta/2} e^{+ \pi\eta /2}
K_{i\eta}\left(\sqrt{w x^+ x^- }\right)\!,
& \qquad x^{\pm} < 0\,,
\\
2 \left(x^+ \over - x^-\right)^{i\eta/2}
K_{i\eta}\left(i \sqrt{w x^+ ( - x^- ) }\right)\!,
& \qquad x^+ > 0 > x^-\!,
\\
2 \left( - x^+ \over x^-\right)^{i\eta/2}
K_{i\eta}\left( - i \sqrt{w ( - x^+) x^- }\right)\!.
& \qquad x^- > 0 > x^+\!.
\end{array}
\right.
\end{equation}
%

%\ifpreprintsty \pagebreak \fi
Using these results in Eq.\ (\ref{eq:psi+}),
the closed form expression for $\psi^{(+)}$ may then be immediately written
as
\begin{eqnarray}
\lefteqn{\phi^{(+)}_{\eta, n}(x^+, x^-, y, \eta) =
\sqrt{2\pi}\psi^{(+)}_{w, \eta, n}(x^+, x^-, y, \eta) =} \hspace{5em}
\nonumber \\
& & \left\{%
\begin{array}{ll}
\left\{
\left[e^{i{y \over Y_0}\left({n\pi\over 2} + \alpha \eta \right)}
\over i^n ( 2\pi)\sqrt{2 Y_0}\right]
\left(x^+ \over x^-\right)^{i\eta/2} e^{- \pi\eta /2}
\right.
\\
\noalign{\vspace{0.1ex}}
\left.
\null  + \left[e^{- i{y \over Y_0}\left({n\pi\over 2} + \alpha \eta \right)}
\over i^{-n} ( 2\pi)\sqrt{2 Y_0}\right]
\left(x^+ \over x^-\right)^{- i\eta/2} e^{\pi\eta /2}
\right\}
K_{i\eta}\left(\sqrt{w x^+ x^- }\right)\!,
& \qquad x^{\pm} > 0\,,
\\
\noalign{\vspace{0.2ex}}
\left\{
\left[e^{i{y \over Y_0}\left({n\pi\over 2} + \alpha \eta \right)}
\over i^n ( 2\pi)\sqrt{2 Y_0}\right]
\left(- x^+ \over - x^-\right)^{i\eta/2} e^{+ \pi\eta /2}
\right.
\\
\noalign{\vspace{0.1ex}}
\left.
\null + \left[e^{- i{y \over Y_0}\left({n\pi\over 2} + \alpha \eta \right)}
\over i^{-n} ( 2\pi)\sqrt{2 Y_0}\right]
\left( - x^+ \over  - x^-\right)^{- i\eta/2} e^{ - \pi\eta /2}
\right\}
K_{i\eta}\left(\sqrt{w x^+ x^- }\right]\!,
& \qquad x^{\pm} < 0\,,
\\
\noalign{\vspace{0.2ex}}
\left\{
\left[e^{i{y \over Y_0}\left({n\pi\over 2} + \alpha \eta \right]}
\over i^n ( 2\pi)\sqrt{2 Y_0}\right]
\left(x^+ \over - x^-\right)^{i\eta/2}
\right.
\\
\noalign{\vspace{0.1ex}}
\null + \left.
\left[e^{- i{y \over Y_0}\left({n\pi\over 2} + \alpha \eta \right)}
\over i^{-n} ( 2\pi)\sqrt{2 Y_0}\right]
\left(x^+ \over  - x^-\right)^{- i\eta/2}
\right\}
K_{i\eta}\left(i \sqrt{w x^+ ( - x^- ) }\right]\!,
& \qquad x^+ > 0 > x^-\!,
\\
\noalign{\vspace{0.2ex}}
\left\{
\left[e^{i{y \over Y_0}\left({n\pi\over 2} + \alpha \eta \right)}
\over i^n ( 2\pi)\sqrt{2 Y_0}\right]
\left( - x^+ \over x^-\right)^{i\eta/2}
\right.
\\
\noalign{\vspace{0.1ex}}
\null + \left.
\left[e^{- i{y \over Y_0}\left({n\pi\over 2} + \alpha \eta \right]}
\over i^{-n} ( 2\pi)\sqrt{2 Y_0}\right]
\left( - x^+ \over x^-\right)^{- i\eta/2}
\right\}
K_{i\eta}\left( - i \sqrt{w ( - x^+) x^- }\right)\!,
& \qquad x^- > 0 > x^+\!,
\end{array}
\right.
\end{eqnarray}
where the fact that $K_{i\eta}$ is even in $\eta$ has been used, and
\begin{equation}
w = w(n,\eta) = m^2 +  (n\pi/2 + \alpha\eta)^2/Y_0^2 \,.
\end{equation}
Note that $K_{i\eta}$ is real for real values of its argument.
The wave function has a branch point at $x^{\pm} = 0$, reflecting the causal
anomalies which appear at the Cauchy horizons.
However, wave packets constructed with $\phi^+$ are well behaved
along the horizons.

The wave functions $\phi_{\eta, n}$ defined in Eq.\ (\ref{eq:phi+}) form a
complete set of functions on the spacelike hyperboloids in the past and in the
future regions.  They are orthonormal under integration over the surface in
Eq.\ (\ref{eq:surface-integral}).
In particular, for a spacelike hyperboloid in the past or future region, let
$x^{\pm} = \pm \tau e^{\pm\xi}$, where $\tau$ is positive (negative) in the
future (past) region; the metric becomes
\begin{equation}
d^2s =  -d\tau^2 + \tau^2 d\xi^2 + dy^2\,,
\end{equation}
%
%\ifpreprintsty \pagebreak \fi
and the integral over the spacelike hyperboloid $\tau = \mbox{const}$ is
\begin{eqnarray}
\lefteqn{%
\int_{-\infty}^{\infty}\,d\xi \int_{-Y_0}^{Y_0}\,dy
\tau i \left[
\phi^{(+)}_{\eta_1,n_1}(\tau e^{\xi}, - \tau e^{-\xi}, y)^{\ast}
i {\mathrel{ \mathop{\partial}\limits^{\leftrightarrow} }
\over
\partial\tau}
\phi^{(+)}_{\eta,n}(\tau e^{\xi}, - \tau e^{-\xi}, y)
\right]
 } \hspace{ 5em}
\nonumber \\
& = & \tau\left[K_{i\eta_1}\left(i\tau\sqrt{w'}\right)^{\ast}
i {\mathrel{ \mathop{\partial}\limits^{\leftrightarrow} }
\over
\partial\tau}
K_{i\eta}\left(i\tau\sqrt{w}\right)\right]
\int_{-\infty}^{\infty}\,d\xi \int_{-Y_0}^{Y_0}\,dy
\nonumber \\
& & \left\{
\left[e^{i{y \over Y_0}\left({n_1\pi\over 2} + \alpha \eta_1 \right)}
\over i^{n_1} ( 2\pi)\sqrt{2 Y_0}\right]
e^{i\xi\eta_1} +
\left[e^{- i{y \over Y_0}\left({n_1\pi\over 2} + \alpha \eta_1 \right)}
\over i^{-n_1} ( 2\pi)\sqrt{2 Y_0}\right]
e^{- i\xi\eta_1}
\right\}
\nonumber \\
& & \left\{
\left[e^{i{y \over Y_0}\left({n\pi\over 2} + \alpha \eta \right)}
\over i^n ( 2\pi)\sqrt{2 Y_0}\right]
e^{i\xi\eta} +
\left[e^{- i{y \over Y_0}\left({n\pi\over 2} + \alpha \eta \right)}
\over i^{-n} ( 2\pi)\sqrt{2 Y_0}\right]
e^{- i\xi\eta}
\right\}
\nonumber \\
& = & \delta(\eta_1 - \eta)\delta_{n_1,n}\,,
\label{eq:phi orthonormality}
\end{eqnarray}
where the Wronskian
$z(K_{\nu}(-i z)
i \mathrel{\mathop{\partial}\limits^{\leftrightarrow}}/\partial z
K_{\nu}(iz) ) = \pi$ is used to evaluate the overall factor.
Note that the integral with $\phi^{(+)*}$ replaced by $\phi^{(-)*}$ vanishes
because the Wronskian then vanishes.

In the acausal region, the specification of a surface is more complicated.
Because of the boost, the surface must have its intersection with
the boundaries $y = \pm Y_0$ be continuous; that is, the identified point is
also in the surface.
This can be achieved in a variety of ways, but in no case is the resultant
surface everywhere spacelike.
One particularly simple choice is to let the surface be defined by
\begin{equation}
x^{\pm} = \left\{
\begin{array}{ll}
\xi e^{\pm\tau} e^{\mp {y\alpha/ Y_0}}\,, & \qquad x^{\pm} > 0 \,,
\\
- \xi e^{\mp\tau} e^{\mp {y\alpha/ Y_0}}\,, & \qquad x^{\pm} < 0\,,
\end{array}
\right.
\end{equation}
where $\tau$ is constant.
With this change of variables the metric for the space becomes
\begin{equation}
ds^2 = d\xi^2 + dy^2 - \xi^2(d\tau \mp \alpha dy/Y_0 )^2\,,
\end{equation}
where the $\mp$
sign is negative for the $x^{\pm} > 0$ region, and positive for the
$x^{\pm} < 0$ region.
The normal one-form is $d\tau$, and the normal derivative becomes
\begin{equation}
\vec n_{\pm} = \left(1\over\xi^2\right)
{\vec \partial \over \partial\tau} \mp \left(\alpha\over Y_0\right)
\left[{\vec \partial \over \partial y} \pm \left(\alpha\over Y_0\right)
{\vec \partial \over \partial \tau} \right]\!.
\end{equation}
The integral over the surface $\tau = \mbox{const}$ is
\begin{eqnarray}
\lefteqn{%
I_{n_1,\eta_1;n,\eta} =
\int_{0}^{\infty}\,d\xi \int_{-Y_0}^{Y_0}\,dy \xi }
%\hspace{5em}
\nonumber \\
& \left[
\phi^{(+)}_{\eta_1,n_1}(\xi e^{\tau- \alpha y/Y_0},
\xi e^{-\tau + \alpha}, y)^{\ast}
i \mathrel{ \mathop{n}\limits^{\leftrightarrow} }_+
\phi^{(+)}_{\eta,n}(\xi e^{\tau- \alpha y/Y_0},
\xi e^{-\tau + \alpha y/Y_0}, y) \right. &
\\
&\null + \left.
\phi^{(+)}_{\eta_1,n_1}( - \xi e^{- \tau - \alpha y/Y_0},
- \xi e^{\tau + \alpha}, y)^{\ast}
i \mathrel{ \mathop{n}\limits^{\leftrightarrow} }_-
\phi^{(+)}_{\eta,n}(- \xi e^{- \tau - \alpha y/Y_0},
- \xi e^{\tau + \alpha y/Y_0}, y)
\right]\!, &
\nonumber
\end{eqnarray}
where
\begin{eqnarray}
\lefteqn{%
\phi^{(+)}_{\eta,n}(\xi e^{\tau- \alpha y/Y_0},
\xi e^{-\tau + \alpha y/Y_0}, y)
=}\hspace{5em}
\nonumber \\
& & \left( 1\over 2\pi\right) K_{i\eta}\left(\xi\sqrt{w(n,\eta)}\right)
\left[
e^{- \pi \eta} E_n(y, \tau; \eta)
+
e^{\pi\eta} E_{-n}( y, \tau; -\eta)
\right]\,,
\end{eqnarray}
and
\begin{eqnarray}
\lefteqn{%
\phi^{(+)}_{\eta,n}(- \xi e^{- \tau- \alpha y/Y_0},
- \xi e^{\tau + \alpha y/Y_0}, y) =} \hspace{5em}
\nonumber \\
& & \left( 1\over 2\pi\right) K_{i\eta}\left(\xi\sqrt{w(n,\eta)}\right)
\left[
e^{ \pi \eta} E_n(y, \tau;  - \eta)
+
e^{- \pi\eta} E_{-n}( y, \tau; \eta)
\right]\,,
\end{eqnarray}
with
\begin{equation}
E_n(y, \tau; \eta) \equiv
{
e^{i n\pi y / 2 Y_0} e^{i\eta \tau} \over i^n \sqrt{2 Y_0}
}\!.
\end{equation}
The integral $I_{n_1,\eta_1;n,\eta}$ requires the $y$ integral
\begin{eqnarray}
\lefteqn{%
\int_{-Y_0}^{Y_0}\,dy
\left\{\left[ E_{n_1}(y,\tau;\eta_1)^{\ast}
i \mathrel{ \mathop{n}\limits^{\leftrightarrow} }_+
E_{n}(y,\tau;\eta) \right] +
\left[( n, n_1 ) \rightarrow ( - n, - n_1 )\right]\right\}
} \hspace{5em}
\nonumber \\
& = & \delta_{n_1,n} e^{i\tau(\eta - \eta_1)}
\left[-( \eta^2 - \eta_1^2)/\xi^2 + w( n, \eta) - w(n, \eta_1)\right]
/(\eta - \eta_1)\,,
\end{eqnarray}
and
\begin{eqnarray}
\lefteqn{%
\left[-( \eta^2 - \eta_1^2)/\xi^2 + w( n, \eta) - w(n, \eta_1)\right]
K_{i\eta_1}\left(\xi\sqrt{w(n,\eta_1)}\right)
K_{i\eta}\left(\xi\sqrt{w(n,\eta)}\right)
} \hspace{10em}
\nonumber \\
& = & { 1 \over \xi }
{\partial \over \partial \xi} \xi
\left[
K_{i\eta_1}\left(\xi\sqrt{w(n,\eta_1)}\right)
{\mathrel{\mathop{\partial}\limits^{\leftrightarrow}}
\over
\partial \xi}
K_{i\eta}\left(\xi\sqrt{w(n,\eta)}\right)
\right]\!.
\label{eq:xi-integrand}
\end{eqnarray}
The integral over $\xi$ of this result would vanish if it were not for the
singular behavior as $\xi \rightarrow 0$ and the pole at $\eta = \eta_1$.
The term which gives the singularity is
\begin{eqnarray}
\lefteqn{%
\int_0\,\left[d\xi  \over \xi\right]
K_{i\eta_1}\left(\xi a'\right) K_{i\eta}\left(\xi a\right)
}
\hspace{5em} \nonumber \\
& \simeq &
\int_0\,d\xi \left(\eta + \eta_1 \over 4 \xi\right)
\left[\Gamma(i\eta_1) (\xi a'/2)^{i\eta_1}
+\Gamma( - i\eta_1) (\xi a'/2)^{ - i\eta_1} \right]
\nonumber \\
& &
\left[\Gamma(i\eta) (\xi a/2)^{i\eta}
+\Gamma( - i\eta) (\xi a/2)^{ - i\eta} \right]
\nonumber \\
&\simeq &\left\{ \left[  \left( - i \Gamma( i\eta) \Gamma( - i \eta_1)
\over 4 (\eta - \eta_1 - i\epsilon )  \right) + ( \eta_1 \rightarrow - \eta_1)
\right]
+ \left[ ( \eta \rightarrow - \eta )\right] \right\}
\nonumber \\
& = &\left[ \pi \delta( \eta - \eta_1)
\Gamma( i\eta) \Gamma( - i \eta)
\over 2   \right] + ( \eta_1 \rightarrow - \eta_1)
\nonumber \\
& = &\left[ \left(  \pi^2 \delta( \eta - \eta_1)
\over \eta ( e^{\pi\eta} - e^{-\pi\eta} )   \right) +
( \eta_1 \rightarrow - \eta_1) \right]\!.
\end{eqnarray}
These results may be combined to show that
\begin{equation}
I_{n_1,\eta_1;n,\eta} = \delta_{n_1,n}\delta(\eta_1 - \eta)\,.
\end{equation}

These results imply that the reduction formula can be written in the form
\begin{equation}
\int\,d\sigma_{\mu} \langle 0 | \left[\phi^{(+)}_{\eta, n}(x)^*
i \mathrel{ \mathop{\partial^{\mu}}\limits^{\leftrightarrow} } \phi(x) \right]
= \langle \eta, n ; \sigma |\,,
\label{eq:reduction}
\end{equation}
where the $\sigma$ in the specification of the state denotes the surface over
which the integral is done.
In the case of the free field, the result is independent of the surface, but in
the case of interactions the result depends upon the surface.
These results also imply particle conservation for the free theory.
The number of particles which reach the final surface is the same regardless
of whether or not they traverse the acausal region between the initial and
final
surfaces.

\appendix{Images}
\label{Images}

In order to evaluate the sums over modes which appear at various places,
the following expression must be evaluated
\begin{equation}
I(y, a, \eta) = \sum_{n = -\infty}^{\infty}
{e^{i(n\pi/2 + \alpha\eta)(y / Y_0)}
\over
a + (n\pi/2 + \alpha\eta)^2/Y_0^2 } \,,
\end{equation}
where $ 0 < y < 4 Y_0$.
It may be rewritten as
\begin{equation}
I(y, a, \eta) = \oint_C dz
{e^{i(z\pi/2 + \alpha\eta)(y / Y_0)} \over e^{2\pi i z} - 1}
{ 1 \over a + (z\pi/2 + \alpha\eta)^2/Y_0^2 } \,,
\end{equation}
where the contour $C$ encloses all the integers along the real axis, but does
not enclose the zeros of the second denominator.
Because of the bounds $0 <  y < 4Y_0$,
the integrand goes to zero exponentially as
$\mbox{Im $z$} \rightarrow \pm\infty$, and the contour can be opened out
to infinity, picking up the poles at the zeros of the second denominator.
This evaluation yields
\begin{eqnarray}
I(y, a, \eta) & = &  \left( 2 Y_0 \over \sqrt{a} \right)
\left[
{e^{ - y \sqrt{a}} \over  1 - e^{ - 4 Y_0 \sqrt{a}}e^{ -i 4\alpha\eta} } +
{e^{ y \sqrt{a}} \over e^{ 4 Y_0 \sqrt{a}}e^{ -i 4\alpha\eta} - 1}
\right]
\nonumber \\
& = & \sum_{ n_1 = 0}^{\infty}\left(2 Y_0 \over \sqrt{a} \right)
\left[e^{- ( y + 4n_1 Y_0)\sqrt{a} } e^{-i4 n_1\alpha\eta} +
e^{( y - 4(n_1+1) Y_0)\sqrt{a} } e^{+i4 (n_1+1)\alpha\eta}\right]
\\
& = & \sum_{ n_1 = -\infty}^{\infty}\left(2 Y_0 \over \sqrt{a} \right)
e^{- |\, y + 4n_1 Y_0\,|\sqrt{a} }e^{-i4 n_1\alpha\eta} \,,
\nonumber
\end{eqnarray}
where $\mbox{ Re $\sqrt{a}$} > 0$.
If $- 4Y_0 < y < 0$, the same result is obtained by noting
that $I(y,a,\eta) = I(-y, a, -\eta)$ and replacing the summation index
$n_1$ by $ -n_1 $; thus the result holds for $ -4Y_0 < y < 4 Y_0$.

Similarly, the sum
\begin{equation}
I^-(y, a, \eta) = \sum_{n = -\infty}^{\infty}
{e^{i(y / Y_0)(n\pi/2 + \alpha\eta)} ( -1 )^n
\over
a + (n\pi/2 + \alpha\eta)^2/Y_0^2 } \,,
\end{equation}
where $ 0 < y < 4 Y_0$, may be rewritten as
\begin{equation}
I^-(y, a, \eta) = \oint_C dz
{e^{i((y + 2 Y_0) / Y_0)(z\pi/2 + \alpha\eta)} e^{-i2\alpha\eta}
\over
e^{2\pi i z} - 1}
{ 1 \over a + (z\pi/2 + \alpha\eta)^2/Y_0^2 } \,.
\end{equation}
Then,
\begin{eqnarray}
I^-(y, a, \eta) & = &  \left( 2 Y_0 \over \sqrt{a} \right)
\sum_{ n_1 = -\infty}^{\infty}
e^{- |\, y + (4n_1 + 2) Y_0\,|\sqrt{a} }e^{-i(4n_1 + 2)\alpha\eta}\,,
\nonumber
\end{eqnarray}
where $\mbox{ Re $\sqrt{a}$} > 0$, and $ -4Y_0 < y + 2 Y_0 < 4 Y_0$.

The sum over the $\psi^{(\pm)}$ modes in the expression for the Green's
function may now be expressed as
\begin{eqnarray}
\lefteqn{\int_0^{\infty}\,d\eta \sum_{n = -\infty}^{\infty}
\left[
\psi^{(+)}_{\lambda,\eta,n}(x')^{\ast}
\psi^{(+)}_{\lambda,\eta,n}(x) +
\psi^{(-)}_{\lambda,\eta,n}(x')^{\ast}
\psi^{(-)}_{\lambda,\eta,n}(x) \right] } \hspace{ 5em}
\nonumber \\
& = &{1 \over 4 ( 2\pi)^3 }
\int_0^{\infty}\,d\eta \int_{-\infty}^{\infty}\,dk^+dk^-dk'^+dk'^-
\delta(\lambda + k^+k^-) \delta(k^+k^- - k'^+k'^-)
\nonumber \\
& & \sum_{n = -\infty}^{\infty}
{ e^{i(kx -k'x')} \over
k^+k^- + m^2 -i \epsilon + ( n\pi/2 + \alpha \eta)^2/Y_0^2}
\nonumber \\
& & \left\{
{e^{i(( y- y')/Y_0)(n \pi/2 + \alpha \eta )} \over 2 Y_0}
\left[ k^+(-k'^-) \over (-k^-)k'^+ \right]^{i\eta/2} +\mbox{c.c.} \right.
\nonumber \\
& & \left. \null +
{e^{i(( y + y')/Y_0)(n \pi/2 + \alpha \eta )}(-1)^n \over 2 Y_0}
\left[ k^+k'^+ \over (-k^-)( -k'^-) \right]^{i\eta/2} +\mbox{c.c.}
\right\}
\nonumber \\
& = &{1 \over 4 ( 2\pi)^3 }
\int_{-\infty}^{\infty}\,d\eta \int\,dk^+dk^-dk'^+dk'^-
\delta(\lambda + k^+k^-) \delta(k^+k^- - k'^+k'^-)
\nonumber \\
& & \sum_{n_1 = -\infty}^{\infty}
{ e^{i(kx -k'x')} \over \sqrt{a} }
\left\{ e^{- | y - y' + 4n_1Y_0 |\sqrt{a} }
\left[(e^{-4n_1\alpha}k^+)(-k'^-) \over (- e^{4n_1\alpha}k^-)k'^+)
\right]^{i\eta/2}
\right.
\nonumber \\
& & \left. \null +
e^{ - | y + y' +(4n_1+2)Y_0 |\sqrt{a}}
\left[(e^{-(4n_1 + 2)\alpha}k^+)k'^+ \over (- e^{(4n_1 + 2)\alpha}k^-)(- k'^-)
\right]^{i\eta/2}
\right\} \,,
\end{eqnarray}
where $ a = k^+k^- + m^2 - i \epsilon$, $\mbox{Re} \sqrt{a} > 0$, and
$kx = (k^+x^- + k^-x^+)/2$.
The $k^{\pm}$ integration variables can then be scaled by $e^{\mp4\alpha}$;
the latter factors then appear multiplying $x^{\pm}$, and,
using the image variables defined in Eq.\ (\ref{eq:images}) and
letting $k^{\pm} \rightarrow - k^{\mp}$ in the second set of terms,
the expression may be rewritten as
\begin{eqnarray}
\lefteqn{\int_0^{\infty}\,d\eta \sum_{n = -\infty}^{\infty}
\left[
\psi^{(+)}_{\lambda,\eta,n}(x')^{\ast}
\psi^{(+)}_{\lambda,\eta,n}(x) +
\psi^{(-)}_{\lambda,\eta,n}(x')^{\ast}
\psi^{(-)}_{\lambda,\eta,n}(x) \right] } \hspace{ 5em}
\nonumber \\
& = &{1 \over 4 ( 2\pi)^3 }
\int_{-\infty}^{\infty}\,d\eta dk^+dk^-dk'^+dk'^-
\delta(\lambda + k^+k^-) \delta(k^+k^- - k'^+k'^-)
\nonumber \\
& & \sum_{n_1 = -\infty}^{\infty}
{ e^{i(kx_{n_1} -k'x')} \over \sqrt{a} }
e^{- | y_{n_1} - y' |\sqrt{a} }
\left[k^+(-k'^-) \over (- k^-)k'^+ \right]^{i\eta/2}\!.
\end{eqnarray}
The $\eta$ and $k'$ integrals can be done directly yielding
\begin{eqnarray}
\lefteqn{\int_0^{\infty}\,d\eta \sum_{n = -\infty}^{\infty}
\left[
\psi^{(+)}_{\lambda,\eta,n}(x')^{\ast}
\psi^{(+)}_{\lambda,\eta,n}(x) +
\psi^{(-)}_{\lambda,\eta,n}(x')^{\ast}
\psi^{(-)}_{\lambda,\eta,n}(x) \right] } \hspace{ 5em}
\nonumber \\
& = & {1 \over 4 ( 2\pi)^2 }
\sum_{n_1 = -\infty}^{\infty}
\int_{-\infty}^{\infty}\,dk^+dk^- \delta(\lambda + k^+k^-)
{ e^{ik(x_{n_1} -x')} \over \sqrt{a} }
e^{- | y_{n_1} - y' |\sqrt{a} } \,,
\end{eqnarray}
where $\lambda > 0$, and a similar argument yields
\begin{eqnarray}
\lefteqn{\int_{-\infty}^{\infty}\,d\eta \sum_{n = -\infty}^{\infty}
\psi^{s}_{\lambda,\eta,n}(x')^{\ast}
\psi^{s}_{\lambda,\eta,n}(x) } \hspace{ 5em}
\nonumber \\
& = & {1 \over 4 ( 2\pi)^2 }\sum_{n_1 = -\infty}^{\infty}
\int_{-\infty}^{\infty}\,dk^+dk^- \delta(\lambda + k^+k^-)
{ e^{ik(x_{n_1} -x')} \over \sqrt{a} }
e^{- | y_{n_1} - y' |\sqrt{a} } \,,
\end{eqnarray}
where $\lambda < 0$.

The familiar result for the Green's function in Minkowski spacetime is given
by
\begin{eqnarray}
G_M(x,x')
& = & \int\,{d^3k \over (2\pi)^3}
{e^{ik\cdot(x - x')} \over k^2 + m^2 -i\epsilon }
\nonumber \\
& = & {1 \over 4 (2\pi)^2 }
\int_{-\infty}^{\infty}\,{dk^+ dk^- \over  \sqrt{m^2 + k^+k^-}}
e^{ik(x - x')} e^{- |y - y' | \sqrt{m^2 + k^+k^-} } \,,
\end{eqnarray}
where $k\cdot x = kx + k_y y$.
When these results are included in Eq.\ (\ref{eq:green's function:G}),
the full Green's function becomes
\begin{equation}
G(x,x') = \sum_{n = -\infty}^{\infty} G_0(x_n, x')
 = \sum_{n = -\infty}^{\infty} G_0(x, x'_n)\,,
\end{equation}
where the $x'_n$ are the images given by Eq.\ \ref{eq:images}.
%%
%\begin{equation}
%\begin{array}{rcl}
%x^{\pm}_n & = & \left\{
%		    \begin{array}{rl}
%		    e^{ \mp 2 n \alpha} x^{\pm}\,, &\mbox{\quad$n$ even,} \\
%		    - e^{ \mp 2 n \alpha} x^{\mp}\,, &\mbox{\quad$n$ odd,}
%		    \end{array}
%		\right. \\
%	\multicolumn{3}{c}{}\\
%	y_n & = & \left\{
%		    \begin{array}{rl}
%		    y + 2 n Y_0\,, &\mbox{\quad$n$ even,} \\
%		    - y + 2 n Y_0\,, &\mbox{\quad$n$ odd.}
%		    \end{array}
%		\right.
%		\end{array}
%\end{equation}
%%
%
%
%\input{references}

\ifprinter
\else
\begin {figure}[p]%			Figure 1
    \caption
	{%
	\advance\baselineskip by -8pt
	The space for a single point mass of deficit angle $\pi$. The space
	consists of the region above the line with the tick marks denoting
	the identified lines.
	}%
    \label {fig:1mass}
\end {figure}
\begin {figure}[p]%			Figure 2
    \caption
	{%
	\advance\baselineskip by -8pt
	The space for a two point masses of deficit angle $\pi$. The space
	consists of the region between the lines with the single, and double
	tick marks respectively denoting the identified lines.
	}%
    \label {fig:2mass}
\end {figure}
\begin {figure}[p]%			Figure 3
    \caption
	{%
	\advance\baselineskip by -8pt
	The hyperbolae show the surfaces on which images of a point in the
	physical region lie.
	The $y$ axis is perpendicular to the graph, and the physical space
	consists of the region $-Y_0 < y < Y_0$.
	}%
    \label {fig:hyperbolae}
\end {figure}
\begin {figure}[p]%			Figure 4
    \caption
	{%
	\advance\baselineskip by -8pt
	The left and right figures respectively show closed future directed
	timelike curves of winding number 1 and 2.
	The identified points are labeled in increasing
	temporal order from A to B (D) along the
	respective curves.
	}%
    \label {fig:ctcs}

\end {figure}
\fi
\end{document}